\documentclass[aps,prd,10pt,twocolumn,nofootinbib,superscriptaddress,floatfix]{revtex4-2}

\usepackage{amsmath,amssymb,amsfonts,bm}
\usepackage{graphicx}
\usepackage{booktabs}
\usepackage{microtype}
\usepackage{hyperref}
\newcommand{\ket}[1]{\left|#1\right\rangle}
\usepackage{tikz}
\usetikzlibrary{calc, positioning, decorations.pathmorphing, shapes.geometric, arrows.meta}

\hypersetup{colorlinks=true,linkcolor=blue,citecolor=blue,urlcolor=blue}

\begin{document}

\title{Axion polarimetry with parametric readout: photon-number conversion and sensitivity limits}

\author{A. Capolupo}
\email{capolupo@sa.infn.it}
\affiliation{Dipartimento di Fisica ``E.R. Caianiello'', Università di Salerno,
	and INFN -- Gruppo Collegato di Salerno, Via Giovanni Paolo II, 132,
	84084 Fisciano (SA), Italy}

\author{G. Pisacane}
\email{gpisacane@unisa.it}
\affiliation{Dipartimento di Fisica ``E.R. Caianiello'', Università di Salerno,
	and INFN -- Gruppo Collegato di Salerno, Via Giovanni Paolo II, 132,
	84084 Fisciano (SA), Italy}

\author{R. Serao}
\email{r.serao@ssmeridionale.it}
\affiliation{Scuola Superiore Meridionale, Largo San Marcellino 10, 80138 Napoli, Italy
	Istituto Nazionale di Fisica Nucleare, Sezione di Napoli,
	Complesso Universitario di Monte S. Angelo, Via Cintia Edificio 6, 80126 Napoli, Italy}

\begin{abstract}

We propose a bias-assisted active photon-number conversion scheme for
axion-induced polarization rotation in a four-port optical ring cavity
with parametric readout. A calibrated real polarization rotation applied
\emph{before} the final optical parametric oscillator (OPO) converts the
otherwise quadratic response of the summed output photon number into a
linear signal. We derive the resulting photon moments and the extracted-field
covariance, retaining the four-port optical transfer, finite axion
coherence and linewidth. Our principal phenomenological result is a
prospective low-mass sensitivity window at
$m_a=10^{-18}$--$10^{-16}\,\mathrm{eV}$, extending below the
published exclusion bounds included in our comparison, notably the
NuSTAR M82 limits. This is a conditional noise-equivalent projection,
using the adopted $A/\sigma_A=1$ stochastic-power criterion and a
parameterized technical-noise budget, not a confidence-level exclusion
forecast. As a separate readout result, OPO-based total counting reduces
the noise-equivalent coupling by approximately $5\%$ relative to matched
selective attenuation in a favorable higher-frequency band; $5\%$
internal loss together with a $0.10$-rad pump-phase offset leaves a
$3.5\%$ advantage in the stated example. A coherent input with only
the final OPO retains this high-frequency benefit, whereas optimized
coherent polarimetry remains more sensitive in the baseline. Our
results identify both the operating principle of active number conversion
and the stability requirements for a future implementation.

\end{abstract}

\maketitle
\section{Introduction}

Cosmological and astrophysical observations provide evidence
for a substantial nonbaryonic dark-matter component, ranging
from galactic rotation curves and galaxy clusters to the
cosmic microwave background
\cite{Dm1,Dm2,Dm3,Dm4,Dm5,Planck2020,BertoneHooper2018}.
Additional information comes from galactic substructure and
local dynamical measurements \cite{Dm17,deSalasWidmark2021}.
A broad range of particle candidates and detection strategies
has been explored
\cite{BertoneHooperSilk2005,Dm11,SSDM,Dm12,Dm13,Dm14,Dm15,Dm18},
alongside alternative approaches based on modified gravity
\cite{Dm7}.

Among the proposed candidates, axions and axion-like particles
are particularly well motivated as light bosonic dark matter
\cite{Arvanitaki2010}.
The QCD axion also provides a dynamical solution to the
strong-CP problem
\cite{PecceiQuinn1977,PecceiQuinnPRD1977,Weinberg1978,Wilczek1978}.
More generally, light scalar fields can affect neutrino
oscillations \cite{Dm16}, while axion-like cosmological fields
have been investigated through pulsar timing and galactic
dynamics \cite{Axion6,Axion9}.
Other theoretical approaches involve neutrino flavor
condensates and wave-dark-matter halos
\cite{principal,CND14,CND15}.
Further dark-sector scenarios include mirror matter, which
motivates neutron--hidden-neutron interferometric searches
\cite{MM1,MM2,MM3,MM4,MM5,MM6,MM7,Pisacane2},
and light-boson interpretations of anomalous nuclear
transitions and related observables
\cite{R1,R2,R3,R4,R5}.

In the present work, we focus on the electromagnetic coupling
of an ultralight axion field. When sufficiently light and
highly occupied, such a field can be treated as a classical
background on laboratory scales. Its interaction with photons
is described by
\(\mathcal L_{a\gamma}=-g_{a\gamma}aF_{\mu\nu}
\widetilde F^{\mu\nu}/4=g_{a\gamma}a\mathbf E\cdot\mathbf B\),
where $a$ denotes the axion field and $g_{a\gamma}$ its
electromagnetic coupling. A time-dependent background modifies
the propagation phases of opposite photon helicities, producing
a small oscillatory rotation in a linear polarization basis.
This provides the basis for an optical search without an
externally applied magnetic field.

Cavity experiments such as DANCE, LIDA and the Axion Dark-Matter
Birefringent Cavity (ADBC) exploit this polarization conversion
\cite{Obata2018,DANCEAct1,LIDA2024,LiuADBC2019,PandeyADBC2024}.
Their response depends on segment lengths, mirror polarization maps,
mode splitting and extraction ports
\cite{GarciaCely2025,Takidera2025,Kuramoto2025,APE2026}.
A fixed circulation direction alone does not determine whether the
conversion accumulates. We therefore specify a planar four-mirror
geometry and transport both carrier and signal to the detection
ports, retaining the two optical sidebands that determine the
measured quadrature.

The optical sequence considered here uses a coherent $x$-polarized
seed and vacuum in $y$. A first degenerate optical parametric
oscillator (OPO) amplifies the carrier quadrature before the cavity;
a phase-matched inverse OPO subsequently acts on the extracted $x$
polarization \cite{Caves1981}. Two spatially separated squeezing
operations were previously proposed to detect a scalar dark-matter
field \cite{CapolupoScalar2026}. The present pseudoscalar interaction
instead encodes the signal as conversion between orthogonal
polarizations. The different interaction generator leads to
distinct signal and noise properties. Magnetic-field-assisted
axion--photon mixing and geometric-phase probes use a different
experimental setting
\cite{CapolupoAxionQFT2019,CapolupoGeometric2015}.
Axion-mediated fermion interactions and neutron interferometry offer
other complementary probes \cite{Axion5,Axion7}.
It also distinguishes our probe from signal-polarization squeezed-vacuum injection, used by
Martynov and Miao to reduce the detected quadrature noise
\cite{MartynovMiao2020}: the $y$ input to the interaction remains
vacuum in the configuration studied here.

We investigate two physically distinct readout observables. In the
first, corresponding to the conventional polarimetric configuration, a
downstream electro-optic modulator (EOM) produces a reference field in
the selected orthogonal polarization, and the resulting interference
term is linear in the axion-induced rotation. This term cancels,
however, if the photon numbers of both polarizations are simply added.
In the second, which forms our central new construction, we instead
place a \emph{real, calibrated polarization bias before the inverse OPO}
and detect both outgoing polarizations. The active transformation then
operates away from its number-conversion null, producing a first-order
change in their \emph{summed} photon number. The bias can be implemented
as opposite phases of the circular polarizations; a phase delay on one
linear polarization alone is not equivalent. The resulting signal is an
example of pump-assisted photon-number transduction, not photon creation
by the off-resonant axion interaction. This placement and observable
distinguish the proposal from an ordinary downstream-EOM interference
measurement.

We derive the number response and noise covariance for the complete
squeeze--rotate--unsqueeze sequence and evaluate the extracted optical
fields using an explicitly specified four-port cavity. For the first
observable, our principal phenomenological result is that the
\emph{polarimetric} projection, under the chosen optical resources,
local dark-matter density and noise assumptions, potentially probes the
mass interval $10^{-18}$--$10^{-16}\,\mathrm{eV}$ at couplings below the
bounds included in our comparison, notably the NuSTAR M82 limits
\cite{NingSafdi2025}. The result is a noise-equivalent sensitivity
projection under the $A/\sigma_A=1$ convention, where
$A=g_{a\gamma}^2$, rather than a 95\% confidence-level exclusion.
An experimental realization would have to establish the required
noise spectra in the approximately $0.24$--$24\,\mathrm{mHz}$ band.

Turning to the second observable, a separate result concerns
\emph{total-count} readout in a favorable higher-frequency optical band:
at matched optical resources, the final OPO can reduce the
noise-equivalent coupling by approximately $5\%$ relative to a specified
passive selective-attenuation comparator. We test its robustness against
internal loss and pump-phase error and show that the final OPO, rather
than the first squeezing stage, is responsible for this benefit. These
findings do not imply a corresponding advantage for the low-mass
polarimetric projection; in fact, the modeled polarimetric OPO
configuration is slightly less sensitive than the matched passive
comparator. The information bound, exact photon accounting and
output-covariance analysis allow both statements to be compared without
conflating their observables.

We introduce a downstream-EOM polarimetric arrangement as part of our proposed experimental scheme. The biased total-count variant and the
changes required at detection are specified in
Sec.~\ref{sec:bias_total_counts}. Section~\ref{sec:hamiltonian_rwa}
derives the polarization interaction, and
Secs.~\ref{sec:su11_direct_readout}--\ref{sec:ring_cavity} develop
the optical transformations, polarimetric readout and cavity response.
Sections~\ref{sec:power_accounting}--\ref{sec:noise_equivalent_coupling}
give the resource comparison, sensitivity and noise requirements;
Sec.~\ref{sec:comparison_realistic} compares optical searches.
Section~\ref{sec:bias_total_counts} then derives and assesses the
linear total-count readout. Canonical and input--output details are
collected in the appendices. We use $\hbar=c=1$ unless SI units are
explicitly restored.

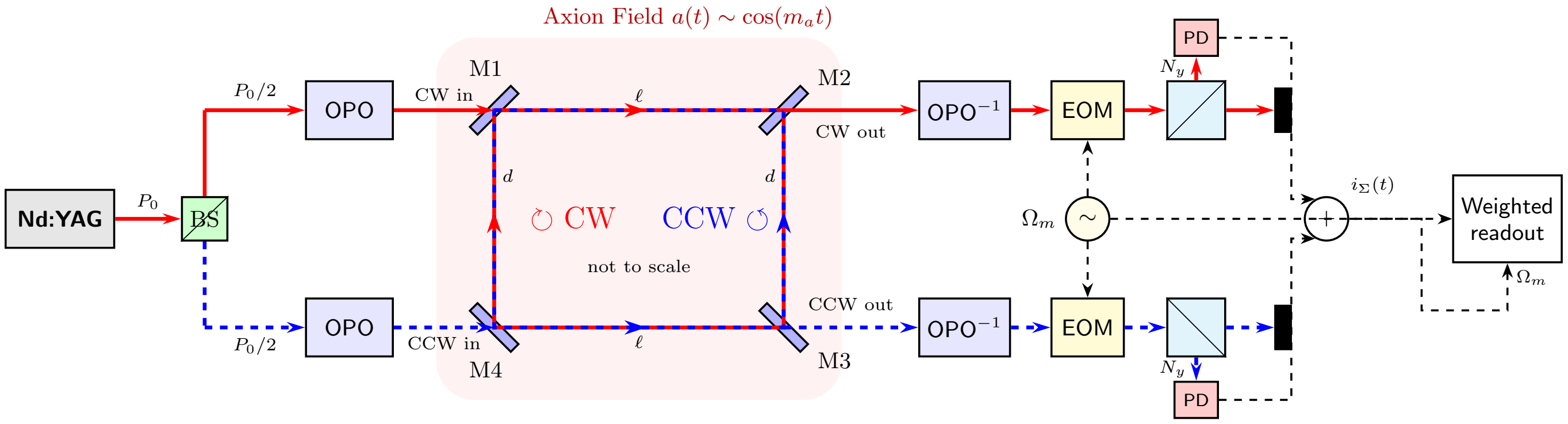
\begin{figure*}[htbp]
		\centering
		\resizebox{\textwidth}{!}{%
			\begin{tikzpicture}[
				laser/.style={rectangle, draw=black, thick, fill=gray!20, minimum width=1.5cm, minimum height=0.8cm, font=\sffamily\bfseries},
				opo/.style={rectangle, draw=black, thick, fill=blue!10, minimum width=1.2cm, minimum height=0.8cm, font=\sffamily},
				mirror/.style={rectangle, draw=black, thick, fill=blue!30, minimum width=0.15cm, minimum height=0.8cm, inner sep=0pt},
				bs/.style={rectangle, draw=black, thick, fill=green!20, minimum width=0.6cm, minimum height=0.6cm},
				eom/.style={rectangle, draw=black, thick, fill=yellow!20, minimum width=1.0cm, minimum height=0.8cm, font=\sffamily},
				pbs/.style={rectangle, draw=black, thick, fill=cyan!10, minimum width=0.8cm, minimum height=0.8cm},
				pd/.style={rectangle, draw=black, thick, fill=red!20, minimum width=0.6cm, minimum height=0.5cm, font=\sffamily\scriptsize},
				dump/.style={rectangle, draw=black, thick, fill=black, minimum width=0.2cm, minimum height=0.6cm},
				beamCW/.style={draw=red, thick, line width=1.5pt},
				beamCCW/.style={draw=blue, thick, dashed, line width=1.5pt},
				beamarrowCW/.style={draw=red, thick, line width=1.5pt, -{Stealth[length=2.5mm, width=1.5mm]}},
				beamarrowCCW/.style={draw=blue, thick, dashed, line width=1.5pt, -{Stealth[length=2.5mm, width=1.5mm]}},
				elec/.style={draw=black, thick, dashed, -{Stealth[length=2mm, width=1.5mm]}}
				]
				%
				\node[laser] (laser) at (0, 0) {Nd:YAG};
				\node[bs] (bs) at (2, 0) {BS};
				\draw (bs.south west) -- (bs.north east);
				
				\draw[beamarrowCW] (laser.east) -- (bs.west) node[midway, above, font=\scriptsize, text=black] {$P_0$};
				
				\node[opo] (opo1_cw) at (4, 1.5) {OPO};
				\node[opo] (opo1_ccw) at (4, -1.5) {OPO};
				
				\draw[beamCW] (bs.north) -- (2, 1.5);
				\draw[beamarrowCW] (2, 1.5) -- (opo1_cw.west) node[midway, above, font=\scriptsize, text=black] {$P_0/2$};
				
				\draw[beamCCW] (bs.south) -- (2, -1.5);
				\draw[beamarrowCCW] (2, -1.5) -- (opo1_ccw.west) node[midway, below, font=\scriptsize, text=black] {$P_0/2$};
				
				\fill[red!5, rounded corners=15pt] (5.2, -2.5) rectangle (10.8, 2.5);
				\node[text=red!70!black, font=\small, anchor=south east] at (10.8, 2.5) {Axion Field $a(t) \sim \cos(m_a t)$};
				\node[mirror, rotate=-45] (m1) at (6, 1.5) {};   \node[above left] at (m1.north west) {M1};
				\node[mirror, rotate=-45] (m2) at (10, 1.5) {};  \node[above right] at (m2.north east) {M2};
				\node[mirror, rotate=45] (m3) at (10, -1.5) {};  \node[below right] at (m3.south east) {M3};
				\node[mirror, rotate=45] (m4) at (6, -1.5) {};   \node[below left] at (m4.south west) {M4};
				\draw[beamarrowCW] (opo1_cw.east) -- (m1.center) node[midway, above, font=\scriptsize, text=black] {CW in};
				\draw[beamarrowCCW] (opo1_ccw.east) -- (m4.center) node[midway, below, font=\scriptsize, text=black] {CCW in};
				\draw[beamCW] (m1.center) -- (m2.center) -- (m3.center) -- (m4.center) -- cycle;
				\draw[beamCCW] (m1.center) -- (m2.center) -- (m3.center) -- (m4.center) -- cycle;
				\node[above, font=\scriptsize] at (8,1.5) {$\ell$};
                \node[below, font=\scriptsize] at (8,-1.5) {$\ell$};
                \node[right, font=\scriptsize] at (6,0.6) {$d$};
                \node[left, font=\scriptsize] at (10,0.6) {$d$};
                \node[font=\scriptsize] at (8,-0.65) {not to scale};
                \node[text=red, font=\large, anchor=east] at (7.8, 0) {$\circlearrowright$ CW};
				\node[text=blue, font=\large, anchor=west] at (8.2, 0) {CCW $\circlearrowleft$};
				\draw[-{Stealth[length=3mm, width=2mm]}, red, thick] (8, 1.5) -- (8.1, 1.5);
				\draw[-{Stealth[length=3mm, width=2mm]}, red, thick] (6, 0) -- (6, 0.1);
				\draw[-{Stealth[length=3mm, width=2mm]}, blue, thick] (8, -1.5) -- (8.1, -1.5);
				\draw[-{Stealth[length=3mm, width=2mm]}, blue, thick] (10, 0) -- (10, 0.1);
				%
				\node[opo] (opo2_cw) at (12.5, 1.5) {OPO$^{-1}$};
				\node[opo] (opo2_ccw) at (12.5, -1.5) {OPO$^{-1}$};
				
				\draw[beamarrowCW] (m2.center) -- (opo2_cw.west) node[midway, below=2pt, font=\scriptsize, text=black] {CW out};
				\draw[beamarrowCCW] (m3.center) -- (opo2_ccw.west) node[midway, above=3pt, font=\scriptsize, text=black] {CCW out};
				
				\node[eom] (eom_cw) at (14.2, 1.5) {EOM};
				\node[eom] (eom_ccw) at (14.2, -1.5) {EOM};
				
				\node[circle, draw=black, thick, fill=yellow!10, minimum size=0.6cm, inner sep=0pt] (osc) at (14.2, 0) {$\sim$};
				\node[left] at (osc.west) {$\Omega_m$};
				\draw[elec] (osc.north) -- (eom_cw.south);
				\draw[elec] (osc.south) -- (eom_ccw.north);
				
				\draw[beamarrowCW] (opo2_cw.east) -- (eom_cw.west);
				\draw[beamarrowCCW] (opo2_ccw.east) -- (eom_ccw.west);
				
				\node[pbs] (pbs_cw) at (15.7, 1.5) {};
				\draw (pbs_cw.south west) -- (pbs_cw.north east);
				\node[pbs] (pbs_ccw) at (15.7, -1.5) {};
				\draw (pbs_ccw.north west) -- (pbs_ccw.south east);
				
				\draw[beamarrowCW] (eom_cw.east) -- (pbs_cw.west) node[midway, above, font=\scriptsize, text=black] {};
				\node[dump] (dump_cw) at (16.9, 1.5) {};
				\draw[beamarrowCW] (pbs_cw.east) -- (dump_cw.west);
				\node[pd] (pd_cw) at (15.7, 2.5) {PD};
				\draw[beamarrowCW] (pbs_cw.north) -- (pd_cw.south) node[midway, left, font=\scriptsize] {$N_y$};
				
				\draw[beamarrowCCW] (eom_ccw.east) -- (pbs_ccw.west) node[midway, below, font=\scriptsize, text=black] {};
				\node[dump] (dump_ccw) at (16.9, -1.5) {};
				\draw[beamarrowCCW] (pbs_ccw.east) -- (dump_ccw.west);
				\node[pd] (pd_ccw) at (15.7, -2.5) {PD};
				\draw[beamarrowCCW] (pbs_ccw.south) -- (pd_ccw.north) node[midway, left, font=\scriptsize] {$N_y$};
				
				\node[circle, draw=black, thick, fill=white, minimum size=0.6cm, inner sep=0pt] (sub) at (17.5, 0) {\textbf{$+$}};
				\node[above right, font=\scriptsize] at (sub.north east)
				{$i_\Sigma(t)$};
				
				\node[rectangle, draw=black, thick, fill=white, minimum width=1.5cm, minimum height=1.2cm, align=center, font=\sffamily] (lockin) at (20, 0) {Weighted\\ readout};
				
				\draw[elec] (pd_cw.east) -- ++(1,0) |- (sub.120);
				\draw[elec] (pd_ccw.east) -- ++(1,0) |- (sub.240);
				\draw[elec] (sub.east) -- (lockin.west);
				\coordinate (refroute) at ($(lockin.south)+(0,-0.65)$);
				\draw[elec]
				(osc.east) -- ++(4.3,0)
				|- (refroute)
				-- (lockin.south)
				node[pos=0.7, right, font=\scriptsize] {$\Omega_m$};
			\end{tikzpicture}%
		}

\caption{Optical sequence and port convention for the planar four-mirror
example. The CW path follows M1--M2--M3--M4 and the CCW path
M4--M3--M2--M1, with inputs at M1/M4 and parallel transmitted outputs
at M2/M3. The long and short segments are $\ell$ and $d$, respectively;
the drawing is not to scale. Each channel contains an OPO before the
cavity and an inverse OPO acting only on $x$ afterward. The EOMs have
equal calibrated axes and phases, followed by crossed analyzers and
photodetectors. For the symmetric mirror and port maps specified in
Appendices~\ref{app:cw_ccw_sign} and \ref{app:input_output}, the axion
signals add in $i_\Sigma=i_{\rm CW}+i_{\rm CCW}$ after subtraction of the
known reference background. The difference is a signal-null monitor in
this model; no automatic rejection of common polarization noise is
assumed. General output maps require calibrated channel weights.
This drawing represents the polarimetric readout. The total-count
variant of Sec.~\ref{sec:bias_total_counts} additionally places the
bias before each inverse OPO and detects both polarizing-beam-splitter (PBS) outputs, replacing
the bright-port dumps with photodetectors.}

\label{fig:apparatus_closed_eom}
	\end{figure*}

\section{Axion--photon Hamiltonian in the off-resonant regime}
\label{sec:hamiltonian_rwa}

We first derive the interaction used in the polarimetric analysis.  The starting point is the axion--photon Lagrangian density
\begin{equation}
    \mathcal L
    =
    -\frac14 F_{\mu\nu}F^{\mu\nu}
    -\frac{g_{a\gamma}}{4}aF_{\mu\nu}\widetilde F^{\mu\nu},
    \label{eq:full_lagrangian_main}
\end{equation}
where the axion field is treated semiclassically: the occupation number of the ultralight dark-matter mode is assumed large enough that the laboratory background can be represented by a classical field. Over a time shorter than its coherence time we use the monochromatic form
\(a(t)=a_0\cos(m_at+\theta_a)\), with \(a_0=\sqrt{2\rho_{\rm local}}/m_a\),
where $m_a$ is the axion mass, $\theta_a$ is the local phase, and $\rho_{\rm local}$ is the local dark-matter energy density. In natural units $m_a$ is also the angular frequency; in SI units $\omega_a=m_ac^2/\hbar$. We neglect spatial gradients across the apparatus, $|\boldsymbol\nabla a|\ll |\dot a|$, so the optical system samples a common axion phase at a given laboratory time.  In Coulomb gauge, with $A^0=0$ and $\boldsymbol\nabla\cdot\mathbf A=0$, one has $\mathbf E=-\dot{\mathbf A}$ and $\mathbf B=\boldsymbol\nabla\times\mathbf A$.  Up to a total time derivative, and under boundary conditions for which the associated surface term vanishes, the photon Lagrangian is canonically equivalent to
\(L_\gamma' = \frac12\int d^3x\left(\dot{\mathbf A}^{\,2}-\mathbf B^2\right) +\frac{g_{a\gamma}}{2}\dot a(t) \int d^3x\,\mathbf A\cdot\mathbf B\).
The corresponding Hamiltonian is
\begin{equation}
    \hat H_\gamma'(t)
    =
    \hat H_{\gamma,0}
    -\frac{g_{a\gamma}}{2}\dot a(t)
    \int d^3x\,\hat{\mathbf A}\cdot\hat{\mathbf B}.
    \label{eq:hamiltonian_prime_main}
\end{equation}
Equation~\eqref{eq:hamiltonian_prime_main} is not an additional interaction to be added to the canonical Hamiltonian obtained directly from Eq.~\eqref{eq:full_lagrangian_main}; the two forms are related by a time-dependent canonical transformation.  The explicit derivation is given in Appendix~\ref{app:birefringence}.

To identify which part of Eq.~\eqref{eq:hamiltonian_prime_main} controls optical polarization, we expand the transverse field in helicity eigenmodes $h=\pm1$. Their polarization vectors satisfy
\(i\mathbf k\times\boldsymbol\epsilon_h(\mathbf k) =hk\,\boldsymbol\epsilon_h(\mathbf k)\),
with $k\equiv|\mathbf k|$. The operator $\hat c_{\mathbf kh}$ annihilates a photon of momentum $\mathbf k$ and helicity $h$, and $\hat N_{\mathbf kh}\equiv\hat c_{\mathbf kh}^\dagger\hat c_{\mathbf kh}$ is the corresponding number operator. After normal ordering,
\begin{align}
    \int d^3x:\hat{\mathbf A}\cdot\hat{\mathbf B}:
    ={}&\sum_{h=\pm1}h\int\frac{d^3k}{(2\pi)^3}
    \Bigg[
    \hat N_{\mathbf kh}
    +\frac12\hat c_{\mathbf kh}\hat c_{-\mathbf kh}
    \nonumber\\
    &\hspace{31mm}
    +\frac12\hat c_{\mathbf kh}^{\dagger}
    \hat c_{-\mathbf kh}^{\dagger}
    \Bigg].
    \label{eq:AB_helicity_main}
\end{align}
The first term in Eq.~\eqref{eq:AB_helicity_main} shifts the two helicities without changing photon number. The remaining terms annihilate or create photon pairs with opposite momenta and therefore represent a physically different, non-number-conserving channel. In the interaction picture these pair operators carry phases $e^{\mp2ikt}$; multiplication by the harmonic axion background can make them slowly varying only near the parametric condition $m_a\simeq2k$. The optical regime investigated here is instead far off that resonance and satisfies
\(m_a\ll k\simeq\omega_{\rm opt}\) and \(g_{a\gamma}|\dot a|\ll k\),
so that the pair terms are rapidly oscillating and can be removed by a rotating-wave approximation (RWA). The resulting Hamiltonian is
\begin{equation}
    \hat H_\gamma^{\rm RWA}(t)
    =
    \sum_{h=\pm1}\int\frac{d^3k}{(2\pi)^3}
    \left[k-h\frac{g_{a\gamma}}{2}\dot a(t)\right]
    \hat N_{\mathbf kh}.
    \label{eq:H_RWA_main}
\end{equation}
For a single optical mode, the axion-dependent part reduces to
\begin{equation}
     {
    \hat H_a^{\rm RWA}(t)
    =
    -\frac{g_{a\gamma}}{2}\dot a(t)
    \left(\hat N_+-\hat N_-\right)
    }.
    \label{eq:H_axion_RWA_main}
\end{equation}
This Hamiltonian is diagonal in the circular-polarization basis and generates opposite phases for the two helicities,
\begin{equation}
    \hat c_h(t_f)
    =e^{-ik(t_f-t_i)}e^{ih\Theta_a(t_f,t_i)}\hat c_h(t_i),
    \label{eq:helicity_phase_main}
\end{equation}
where
\(\Theta_a(t_f,t_i) =\frac{g_{a\gamma}}{2}\left[a(t_f)-a(t_i)\right]\).
The free optical phase is common to both helicities and does not affect polarization, whereas their relative phase is $2\Theta_a$.  Using
\(\hat c_+=(\hat c_x+i\hat c_y)/\sqrt2\) and \(\hat c_-=(\hat c_x-i\hat c_y)/\sqrt2\),
Eq.~\eqref{eq:helicity_phase_main} becomes
\begin{equation}
    \begin{pmatrix}
        \hat c_x(t_f)\\
        \hat c_y(t_f)
    \end{pmatrix}
    =e^{-ik(t_f-t_i)}
    \begin{pmatrix}
        \cos\Theta_a&-\sin\Theta_a\\
        \sin\Theta_a&\cos\Theta_a
    \end{pmatrix}
    \begin{pmatrix}
        \hat c_x(t_i)\\
        \hat c_y(t_i)
    \end{pmatrix}.
    \label{eq:linear_rotation_main}
\end{equation}
Equation~\eqref{eq:linear_rotation_main} is the central single-pass observable used below: the opposite helicity phases become an ordinary rotation by $\Theta_a$ in the fixed linear-polarization basis. This transformation is passive and number conserving. It mixes annihilation operators with one another, leaves the photon vacuum invariant, and should not be confused with the non-number-conserving Bogoliubov dynamics generated by the pair terms in Eq.~\eqref{eq:AB_helicity_main}; those terms have been removed by the off-resonant RWA.

For propagation over a time $L$ in natural units, $t_f=t+L$, and the rotation angle used below is therefore
\begin{equation}
    \psi_a(t,L)\equiv\Theta_a(t+L,t)
    =\frac{g_{a\gamma}}{2}\left[a(t+L)-a(t)\right].
    \label{eq:psi_general_main}
\end{equation}
For this harmonic background,
\begin{align}
    \psi_a(t,L)
    ={}&-\frac{g_{a\gamma}}{m_a}\sqrt{2\rho_{\rm local}}
    \sin\left(\frac{m_aL}{2}\right)
    \nonumber\\
    &\times
    \sin\left(m_at+\theta_a+\frac{m_aL}{2}\right).
    \label{eq:psi_exact_main}
\end{align}
In the low-mass limit $m_aL\ll1$, its peak amplitude becomes
\(|\psi_a|_{\rm amp} \simeq \frac{g_{a\gamma}L}{2}\sqrt{2\rho_{\rm local}}\).

\section{Bright-mode squeezing and orthogonal readout}
\label{sec:su11_direct_readout}

We first consider an ideal lossless sequence. The input operators
$\hat a_x\equiv\hat c_x$ and $\hat a_y\equiv\hat c_y$ describe the
linear polarizations of Sec.~\ref{sec:hamiltonian_rwa}. Both OPOs act on
$x$ alone. We call the bright-mode transformation ``closed'' when the
two operations cancel at zero polarization conversion. This property
of the lossless model does not imply inversion of an open cavity;
its port losses are retained in Appendix~\ref{app:input_output}.

For an input state $\ket{\Psi_{\rm in}}=\ket{\alpha}_x\ket{0}_y$, with coherent amplitude $\alpha$ and single-mode mean photon number $n_0=|\alpha|^2$, the ideal sequence is
\begin{equation}
    \ket{\Psi_{\rm out}}
    =S_x^{-1}(r)U_a(L)S_x(r)
    \ket{\alpha}_x\ket{0}_y,
    \label{eq:sequence_closed}
\end{equation}
where
\(S_x(r)=\exp[r(\hat a_x^{\dagger 2}-\hat a_x^2)/2]\)
acts exclusively on the bright polarization. The dimensionless parameter $r\geq0$ sets the squeezing strength; choosing it real fixes the pump-phase convention for the amplified quadrature. A general single-mode squeeze operator also has a phase, which is absorbed here into the selected pump phase and the seed phase introduced below. The unitary $U_a(L)$ denotes the passive polarization rotation of Eq.~\eqref{eq:linear_rotation_main} through angle $\psi_a$. The single-mode quantity $n_0$ counts photons in the coherent seed of this operator model. Later, the continuous-wave calculation uses the input photon flux and the accumulated number $N_0$ instead.

In the Heisenberg picture,
\begin{align}
    S_x^\dagger(r)\hat a_xS_x(r)
    & =\hat a_x\cosh r+\hat a_x^\dagger\sinh r,
    \nonumber\\
    S_x^\dagger(r)\hat a_yS_x(r)
    & =\hat a_y.
\end{align}
Combining the transformations gives the output operators immediately after the inverse OPO and before the EOM,

\begin{align}
    \hat a_{x,{\rm out}}
    ={}&\hat a_x\cos\psi_a
    -\left(\hat a_y\cosh r-\hat a_y^\dagger\sinh r\right)
    \sin\psi_a,
    \nonumber\\
    \hat a_{y,{\rm out}}
    ={}&\hat a_y\cos\psi_a
    +\left(\hat a_x\cosh r+\hat a_x^\dagger\sinh r\right)
    \sin\psi_a.
    \label{eq:output_ops}
\end{align}

At $\psi_a=0$ this map is the identity. For nonzero rotation the $y$
output retains the amplified $x$ quadrature transferred during the
interaction.

Writing the complex seed amplitude as $\alpha=|\alpha|e^{i\vartheta}$, where $\vartheta$ is its optical phase relative to the amplified quadrature, the mean amplitude of the orthogonal output field is
\(\langle\hat a_{y,{\rm out}}\rangle =|\alpha|\sin\psi_a \left(e^r\cos\vartheta+i e^{-r}\sin\vartheta\right)\).
The amplitude gain depends on this relative phase. Choosing the seed phase along the amplified quadrature, $\vartheta=0$, gives
\(\langle\hat a_{y,{\rm out}}\rangle =e^r\alpha\sin\psi_a \simeq e^r\alpha\psi_a\).

This is a gain of the coherent conversion amplitude at fixed seed
power. In the absence of static polarization mixing, the signal
variance at the null operating point remains $V_y=1$. In contrast,
the signal-polarization squeezed vacuum in
Ref.~\cite{MartynovMiao2020} reduces the measured quadrature variance.
Adding that resource here would also be possible for the conventional
reference instrument considered in Sec.~\ref{sec:power_accounting}.

\subsection{Total outgoing photon number}
\label{subsec:total_photons}

Let $\nu=\sinh r$ and $n_S$ denote the photon number in the
interaction mode after the first OPO. For
$\alpha=\sqrt{n_0}e^{i\vartheta}$,
\begin{align}
 n_S&=n_0[\cosh(2r)+\sinh(2r)\cos(2\vartheta)]+\nu^2,
 \label{eq:interaction_photons_exact}\\
 N_{x,{\rm out}}&=n_0\cos^2\psi_a+\nu^2\sin^2\psi_a,
 \nonumber\\
 N_{y,{\rm out}}&=n_S\sin^2\psi_a.
 \label{eq:Nyout_general}
\end{align}
Here $N_{j,{\rm out}}$ denotes the expectation value of the number
operator. Adding both polarizations gives the exact lossless result
\begin{equation}
 N_{{\rm tot,out}}=n_0+(n_S-n_0+\nu^2)\sin^2\psi_a.
 \label{eq:total_photons_general}
\end{equation}
For the phase-matched seed used below, $\vartheta=0$, this becomes
\begin{equation}
 \Delta N_{\rm tot}
 =\big[n_0(e^{2r}-1)+2\sinh^2r\big]\sin^2\psi_a.
 \label{eq:total_photon_excess}
\end{equation}
Thus $r>0$ and $\sin\psi_a\ne0$ produce a positive photon excess
relative to the incident seed. At $\psi_a=0$ the ideal sequence
returns $n_0$ exactly; at $r=0$ the total is independent of rotation.
For a general seed phase the sign follows
Eq.~\eqref{eq:total_photons_general} and need not be positive.

The axion rotation conserves total optical photon number, whereas
the OPO transformations do not. The net work supplied by their pumps
at the carrier frequency $\omega_0$ is $\hbar\omega_0\Delta N_{\rm tot}$ in the degenerate, undepleted-pump
description. The excess is therefore an active transduction of the
rotation, not photon creation by the off-resonant axion interaction.
The $\nu^2$ terms count spontaneous photons in one normalized optical
mode; in continuous operation they must be summed over the accepted
temporal and spectral modes.
These identities apply to the ideal single-mode sequence. They do
not equate the seed power to the power extracted from one port of an
open cavity; that system is treated in Appendix~\ref{app:input_output}.

At zero polarization bias and small $\psi_a$, the total-number
signal is quadratic in $g_{a\gamma}$. If $\psi_a(t)=\psi_{a,0}\cos\Omega_at$ (with $\Omega_a=m_a$), it contains
a constant term and a component at $2\Omega_a$. A passive EOM after
the inverse OPO preserves $\hat N_x+\hat N_y$; its linear
interference terms cancel in that sum. The readout below therefore
retains a selected polarization. Appendix~\ref{app:bright_mode_closure}
gives the total-count variance, and
Sec.~\ref{sec:power_accounting} compares the photon excess at equal
interaction resources.
Section~\ref{sec:bias_total_counts} instead places a calibrated bias
before the inverse OPO, obtaining a linear response in the total
count itself.

\section{Linear lock-in polarimetric readout}
\label{sec:linear_lockin}

A crossed analyzer alone measures the intensity
$N_{y,{\rm out}}\propto\psi_a^2$. To recover a linear response we apply
an external polarization modulation after the inverse OPO,
\(\psi_m(t)=\psi_0\cos(\Omega_m t)\).
Here $\psi_0\ll1$. The small fraction of the residual carrier rotated
into $y$ provides a reference for the axion-generated field. In this
section $\psi_a$ denotes a real conversion amplitude in the measured
quadrature; the cavity and port transport determining that amplitude
are specified in Sec.~\ref{sec:ring_cavity}.

For the ideal sequence, to first order in the rotation,
\begin{align}
 \hat a_{x,{\rm cl}}&\simeq\hat a_x,\nonumber\\
 \hat a_{y,{\rm cl}}&\simeq\hat a_y+\psi_a
       (\hat a_x\cosh r+\hat a_x^\dagger\sinh r),
 \label{eq:closed_first_order_ops}
\end{align}
where ``cl'' denotes the field before the EOM. The EOM adds
$\psi_m\hat a_x$ to the second line. More generally, define
\begin{equation}
 \langle\hat a_{y,{\rm pre-EOM}}\rangle
     =G_{\rm sig}\alpha\psi_a+O(\psi_a^2).
 \label{eq:Gsig_definition}
\end{equation}
The ideal phase-matched value is $G_{\rm sig}=e^r$.
Below, the cavity conversion is factored into $\psi_a$; additional
signal-path loss enters $G_{\rm sig}$. The reference carrier and the
detected variance must be calculated independently in an open system.

Let $P_0$ be the total seed power before the CW/CCW splitter and
$\dot N_0=P_0/(\hbar\omega_0)$ the corresponding photon flux.
An equal splitter sends $\dot N_0/2$ into each channel. If the ideal
inverse sequence restores each carrier to this flux, the photon rate
at analyzer $s$ is
\begin{equation}
 \begin{split}
 \dot N_{y,s}(t)\simeq\frac{\dot N_0}{2}\big[&\psi_m^2(t)
    +2G_{\rm sig}\psi_m(t)\psi_{a,s}(t)\\
    &+G_{\rm sig}^2\psi_{a,s}^2(t)\big].
 \end{split}
 \label{eq:full_crossed_rate}
\end{equation}
The relative sign of $\psi_{a,s}$ is fixed by the complete output
transport. For the symmetric parallel-port example developed below,
$\psi_{a,{\rm CW}}=\psi_{a,{\rm CCW}}=\psi_a$, and the linear terms
add in the sum. Subtracting its known reference background gives
\(\delta\dot N_\Sigma^{(a)}(t) =2G_{\rm sig}\dot N_0\psi_m(t)\psi_a(t)\).
The independent shot noises of the two detectors add in variance.
This normalization also applies to other calibrated combinations with
equal signal magnitude and equal noise; it does not assume common-noise
rejection.

The reference-dominated regime requires
$\psi_0\gg G_{\rm sig}|\psi_a|_{\rm amp}$ and enough detected reference
photons to dominate electronic and dark noise in the analysis band.
Omitting the inverse OPO would multiply the raw linear signal by
another $e^r$, but would also increase the reference-field shot noise.
The resource comparison is made in Sec.~\ref{sec:power_accounting}.

With final detection efficiency $\eta_{\rm ro}$, the ideal total
reference rate and linear signal are
\begin{align}
 \lambda_\Sigma^{(0)}(t)&=\eta_{\rm ro}\dot N_0\psi_m^2(t),\nonumber\\
 \delta\lambda_\Sigma^{(a)}(t)&=
   2\eta_{\rm ro}\dot N_0G_{\rm sig}\psi_m(t)\psi_a(t).
 \label{eq:poisson_rates_lockin}
\end{align}
For coherent analysis time $T$, write
\(\psi_{a,{\rm rms}}^2=T^{-1}\int_0^T\psi_a^2(t)\,dt\) and \(N_0=\dot N_0T\).
The signal-to-noise ratio is
\begin{equation*}
	\mathrm{SNR}_{\rm cl} =\frac{2{\cal C}_mG_{\rm sig}\sqrt{\eta_{\rm ro}N_0} \,\psi_{a,{\rm rms}}}{\sqrt{V_y}}.
\end{equation*}
Here $V_y$ is the variance of the measured quadrature normalized to
vacuum. When this variance is independent of the modulation waveform,
final detection loss gives
\(V_y=1-\eta_{\rm ro}+\eta_{\rm ro}V_y^{\rm pre}\).
Modulation-dependent noise and background require the count-level
weighting below. The complete covariance calculation is given in
Appendix~\ref{app:input_output}.

\subsection{Poisson weighting and modulation efficiency}

Because the EOM reference $\psi_m(t)$ oscillates in time, the resulting reference shot-noise variance varies across each modulation cycle, making uniform or standard cosine demodulation suboptimal near the modulation nodes. To quantify the resulting modulation efficiency ${\cal C}_m$, let $\lambda_s(t;\theta_{\rm sig})$ denote the mean detected photon rate in channel $s$ (CW or CCW) for a small signal-rotation amplitude $\theta_{\rm sig}$ and a fixed dimensionless waveform $f_s(t)$, such that $\psi_{a,s}(t)=\theta_{\rm sig}f_s(t)$. The null-signal rate $\lambda_s^{(0)}(t)\equiv\lambda_s(t;0)$ includes both the EOM-generated reference and any constant detector background $\lambda_{{\rm bg},s}$ (including dark counts); for the ideal balanced configuration considered above, $\lambda_s^{(0)}(t)=\eta_{\rm ro}\dot N_0\psi_m^2(t)/2+\lambda_{{\rm bg},s}$, where $\dot N_0=N_0/T$ is the total seed photon flux over the integration time $T$. In the Poisson limit, the Fisher information for $\theta_{\rm sig}$ is
\({\cal I}_{\theta_{\rm sig}}=
\sum_s\int_0^T
\frac{[\partial_{\theta_{\rm sig}}\lambda_s(t)]^2}
{\lambda_s^{(0)}(t)}\,dt\),
with the derivative evaluated at $\theta_{\rm sig}=0$. Maximizing the information extracted from the photon counts therefore requires weighting each channel proportionally to the local score $\partial_{\theta_{\rm sig}}\lambda_s/\lambda_s^{(0)}|_{\theta_{\rm sig}=0}$, which inversely weights the time-dependent Poisson noise (whereas any non-Poissonian electronic noise requires its own covariance). Assuming scale separation between the fast modulation and a slowly varying axion envelope ($\Omega_m\gg m_a$), we introduce a periodic demodulation weight $u_m(t)$ and define the normalized total background rate $b_0=\sum_s\lambda_{{\rm bg},s}/(\eta_{\rm ro}\dot N_0)$. The modulation efficiency relative to the ideal zero-background Poisson limit then reads
\({\cal C}_m^2[u_m]=
\langle u_m\psi_m\rangle_m^2/
\langle u_m^2(\psi_m^2+b_0)\rangle_m\),
where $\langle\cdot\rangle_m$ denotes an average over many modulation periods and $0\leq{\cal C}_m\leq1$ (for comparable modulation and axion frequencies, ${\cal I}_{\theta_{\rm sig}}$ is evaluated instead with the full time-dependent signal). For balanced channels, the Poisson-optimal weight reduces to $u_m(t)\propto\psi_m(t)/[\psi_m^2(t)+b_0]$; in practice, using a full likelihood or a regularized score avoids dividing measured data by a vanishing modulation near its nodes when $b_0\to 0$. In the zero-background limit, optimal weighting achieves ${\cal C}_m=1$, whereas a conventional cosine weight gives ${\cal C}_{m,\cos}^2=2/3$. For a sinusoidal modulation $\psi_m(t)=\psi_0\cos(\Omega_m t)$ and fractional analyzer power leakage $\epsilon$, in the absence of other backgrounds, one has $b_0=\epsilon$. Defining $b=\epsilon/\psi_0^2$, the modulation efficiencies to leading order in $\epsilon$ and $\psi_0$ become
\({\cal C}_{m,{\rm opt}}^2=1-\sqrt{b/(1+b)}\)
and
\({\cal C}_{m,\cos}^2=2/(3+4b)\).

Inserting the harmonic axion profile of Eq.~\eqref{eq:psi_exact_main} into the linear count-rate response shifts the signal into the sidebands $\Omega_m\pm m_a$, whose low-mass envelope is
\(\left|\frac{\delta\dot N_\Sigma^{(a)}}{\dot N_0}\right|_{\rm env}
\simeq G_{\rm sig}\psi_0g_{a\gamma}L\sqrt{2\rho_{\rm local}}\).
Because of this oscillation at the axion mass $m_a$, a single integration against $\cos(\Omega_m t)$ over the full observation time $T$ would average the signal component toward zero. Instead, ordinary lock-in demodulation followed by low-pass filtering first extracts a baseband record proportional to $\psi_a(t)$, and a subsequent sine--cosine filter estimates the axion component at the trial frequency $m_a$. Over many axion cycles, $\psi_{a,{\rm rms}}=|\psi_a|_{\rm amp}/\sqrt2$, yielding
\begin{equation}
	\mathrm{SNR}_{\rm cl}
	=\frac{\sqrt2{\cal C}_mG_{\rm sig}\sqrt{\eta_{\rm ro}N_0}
		|\psi_a|_{\rm amp}}{\sqrt{V_y}},
	\label{eq:lockin_snr_peak}
\end{equation}
with corresponding noise-equivalent rotations
\begin{align}
	\delta\psi_{{\rm q,rms}}^{\rm cl}
	&=\frac{\sqrt{V_y}}
	{2{\cal C}_mG_{\rm sig}\sqrt{\eta_{\rm ro}N_0}},\nonumber\\
	\delta\psi_{{\rm q,pk}}^{\rm cl}
	&=\sqrt2\,\delta\psi_{{\rm q,rms}}^{\rm cl},
	\label{eq:closed_shot_noise_angle}
\end{align}
where $N_0$ denotes the total seed photon number of both channels throughout.

\section{Geometry-dependent cavity and output response}
\label{sec:ring_cavity}

The cavity enhances the conversion of the bright carrier into the
orthogonal polarization. Its two polarizations may have different
round-trip field-amplitude survival factors $q_x,q_y$, couplings and
resonance frequencies. We
first derive the round-trip source, then specify the approximations
under which it can be represented by a scalar transfer.

\subsection{Ordered propagation and reflection}

Use the Fourier convention $a(t)=\int a_\Omega e^{-i\Omega t}d\Omega/(2\pi)$,
where $a_\Omega$ is the axion Fourier amplitude and the
positive-frequency component has $\Omega=m_a$ for a monochromatic
background in natural units. A segment of flight time $\tau_j$
produces a rotation referred to its exit time,
\begin{equation}
 \delta\theta_j(\Omega)
  =\frac{g_{a\gamma}a_\Omega}{2}(1-e^{i\Omega\tau_j}),
 \qquad J=\begin{pmatrix}0&-1\\1&0\end{pmatrix}.
 \label{eq:segment_rotation}
\end{equation}
If $A_{j-1}$ transports the carrier to the beginning of the segment,
and $D_j$ transports its generated field to the reference section,
the first-order round-trip source is
\begin{equation}
 \delta M(\Omega)=\frac{g_{a\gamma}a_\Omega}{2}
 \sum_j e^{i\Omega t_{>j}}(1-e^{i\Omega\tau_j})D_jJA_{j-1}.
 \label{eq:ordered_roundtrip_source}
\end{equation}
The matrices include mirror phases, losses and transverse-basis
transport; $t_{>j}$ is the propagation time after segment $j$.
The steady carrier is driven by the input port, while the sideband
response contains $[I-e^{i\Omega T_{\rm rt}}M_0]^{-1}$, with $M_0$
the unperturbed round-trip matrix and $T_{\rm rt}=L_{\rm rt}/c$
the round-trip propagation time. Extraction also requires the
output transmission and any source generated on the partial path
to that port. Appendix~\ref{app:input_output} retains these terms.

For an explicit planar example, take alternating lengths
$\ell,d,\ell,d$, with $L_{\rm rt}=2(\ell+d)$ and local right-handed
polarization bases. An ideal reflection is represented, up to a
common phase, by $R=\operatorname{diag}(-1,1)$, so that
$RJR^{-1}=-J$. In the high-finesse, uniform-carrier approximation,
the source at the end of a long side is
\begin{align}
 B(\Omega)={}&(1-e^{iu})-e^{iu}(1-e^{iv})\nonumber\\
 &+e^{i(u+v)}(1-e^{iu})-e^{i(2u+v)}(1-e^{iv})\nonumber\\
 ={}&(1+e^{i(u+v)})(1-2e^{iu}+e^{i(u+v)}),\nonumber\\
 u={}&\Omega\ell/c,\qquad v=\Omega d/c.
 \label{eq:rectangular_source_kernel}
\end{align}
The successive signs follow from mirror transport, not the sense of
circulation. In the short-side limit,
\(B_0(\Omega)=\lim_{d\to0}B(\Omega) =1-e^{i\Omega L_{\rm rt}/c}\),
and $|B_0|/2=|\sin(\Omega L_{\rm rt}/2c)|$.
This reduction also requires the effective paired-mirror map to
preserve the polarization generator between the long segments.
Elongated paths and the mirror maps are thus essential, as in
DANCE and the analysis of Kuramoto et al.
\cite{Obata2018,Kuramoto2025}; ring topology alone does not prevent
cancellation.

\subsection{Scalar limit, bandwidth and nominal zeros}

Let $T_{x,{\rm in}}$ and $T_{y,{\rm out}}$ be the input-carrier and
output-signal power transmissions. For a resonant carrier, a signal
resonance centered on it, and the uniform-carrier approximation, the
output-referred peak conversion is
\begin{align}
 |\psi_a^{\rm cav}|_{\rm amp}
 ={}&\frac{g_{a\gamma}a_0}{2}|B(m_a)|\nonumber\\
 &\times\left|
 \frac{\sqrt{T_{x,{\rm in}}T_{y,{\rm out}}}}
      {(1-q_x)(1-q_ye^{im_aL_{\rm rt}})}\right|.
 \label{eq:psi_cav_transfer}
\end{align}
Here and below phases containing $m_aL$ use natural units. In SI
units they are $\Omega L/c$, with $\Omega=m_ac^2/\hbar$ when $m_a$
denotes a mass. The common practice of quoting $m_a$ in eV refers to
the rest energy. The sine numerator is recovered only with $B=B_0$.
The two denominator factors describe the carrier buildup and signal
storage, respectively. The exact Airy relation is
\begin{align}
 {\cal F}_j&=\frac{\pi}
 {2\arcsin[(1-q_j)/(2\sqrt{q_j})]},\nonumber\\
 q_j&=(\sqrt{1+s_j^2}-s_j)^2,\qquad
 s_j=\sin\frac{\pi}{2{\cal F}_j},\quad j=x,y.
 \label{eq:q_finesse}
\end{align}
At high finesse $1-q_j\simeq\pi/{\cal F}_j$.
The approximation $T_{x,{\rm in}}\simeq1-q_x$ and
$T_{y,{\rm out}}\simeq1-q_y$ is a critical-coupling normalization;
it does not specify the allocation of all ports in Fig.~\ref{fig:apparatus_closed_eom}.

At low frequency, Eq.~\eqref{eq:rectangular_source_kernel} gives
$B=-i\Omega L_g/c+O(\Omega^2)$, with
\(L_g=2(\ell-d)\) and \(\zeta_{\rm geom}=L_g/L_{\rm rt}=(\ell-d)/(\ell+d)\).
For $\ell>d$ and a flat cavity response the effective length is
\(L_{\rm eff}^{xy} =L_g\frac{\sqrt{T_{x,{\rm in}}T_{y,{\rm out}}}} {(1-q_x)(1-q_y)}\).
Under the critical-coupling approximation,
\(L_{\rm eff}^{xy}\simeq \zeta_{\rm geom}\frac{\sqrt{{\cal F}_x{\cal F}_y}}{\pi}L_{\rm rt}\),
and
\(|\psi_a^{\rm cav}|_{\rm amp} \simeq\frac{g_{a\gamma}L_{\rm eff}^{xy}}{2} \sqrt{2\rho_{\rm local}}\).
Thus $\ell=4.95\,\mathrm m$, $d=0.05\,\mathrm m$ gives
$\zeta_{\rm geom}=0.98$. Equal sides cancel this leading accumulated
term; finite open-port corrections can still leave a residual.

The signal-mode field memory and first bandwidth scale are
\(\tau_{{\rm cav},y}\simeq L_{\rm rt}/[c(1-q_y)]\) and \(m_{\rm cav}\sim(1-q_y)/L_{\rm rt}\), respectively.
At higher frequencies, when the round-trip time matches an integer multiple of the axion oscillation period, the monochromatic source in the limiting kernel $B_0$ averages to zero at
\(m_{a,n}L_{\rm rt}=2\pi n\) (\(n=1,2,\ldots\)),
corresponding to $m_{a,n}\simeq n\,1.24\times10^{-7}\,\mathrm{eV}$ for
$L_{\rm rt}=10\,\mathrm m$. In the finite-$d$ kernel of
Eq.~\eqref{eq:rectangular_source_kernel}, odd multiples of the free
spectral range remain exact zeros, whereas even multiples are generally
lifted. While detuning a scalar cavity denominator cannot remove an unchanged numerator zero, modifying the geometric segment weights or accounting for finite open-port extraction can shift or fill these nominal features; near them, one must also retain the finite axion linewidth over the apparatus response (Sec.~\ref{sec:numerical-analysis}).

Equation~\eqref{eq:psi_cav_transfer} describes the $y$ conversion per
incident carrier. It does not determine the extracted $x$ reference
or its covariance. These are derived separately for a fully specified
four-port example in Appendix~\ref{app:input_output}.

\subsection{Transport to the two detectors}

Let $B_s$ be the post-OPO carrier, $m_s(t)$ the complex EOM reference
rotation and $Y_s(t)$ the axion-generated field, all expressed in the
detector basis and in photon-flux amplitude units. Their linear
photocurrent contribution is
\(\delta i_s(t)=2e\eta_s \operatorname{Re}\{[B_sm_s(t)]^*Y_s(t)\}\),
where $e$ is the elementary charge and $\eta_s$ is the photon
detection efficiency of channel $s$.
A simultaneous sign change of reference and signal leaves this
observable invariant. The free-wave sign reversal in
Appendix~\ref{app:cw_ccw_sign} therefore does not determine the
relative photocurrent signs at parallel extraction ports.

For the symmetric mirror maps, equal EOM phases and parallel outputs
specified in that appendix, the calibrated signal transfers obey
\({\cal H}_{\rm CW}={\cal H}_{\rm CCW}\), so that \({\cal H}_{+}=2{\cal H}_{\rm CW}\) and \({\cal H}_{-}=0\);
we therefore use the channel sum for the sensitivity of this baseline example. More generally, to account for asymmetric optics or correlated disturbances, let $\bm z(\Omega)$ collect the two calibrated records and $\bm{\mathcal H}_a$ denote their axion response vector, defining the linear estimator
\(\Sigma(\Omega)=\bm w^\dagger\bm z(\Omega)\) with signal transfer \({\cal H}_\Sigma=\bm w^\dagger\bm{\mathcal H}_a\).
In the absence of an exact noise-nulling constraint, the optimal linear weights follow the inverse noise cross-spectral matrix $C_n$ via $\bm w\propto C_n^{-1}\bm{\mathcal H}_a$. Conversely, completely rejecting a technical disturbance with response vector $\bm h_n$ while preserving the axion signal requires $\bm w^\dagger\bm h_n=0$ and $\bm w^\dagger\bm{\mathcal H}_a\ne0$, which is impossible whenever $\bm h_n$ and $\bm{\mathcal H}_a$ are parallel. In particular, reversing one EOM phase flips the channel's response to optical polarization disturbances alongside the axion signal, and thus cannot by itself establish common-noise rejection.

\section{Optical resources and selective carrier attenuation}
\label{sec:power_accounting}

The seed power, the power delivered to the interaction stage and the
circulating power are distinct resources. For a bright phase-matched
seed, the first OPO gives
\(P_{\rm int}\simeq e^{2r}P_0\),
with the extra energy supplied by its pump. In channel $s$,
\({\cal B}_{x,s}=T_{x,{\rm in},s}/|1-q_{x,s}e^{i\phi_{x,s}}|^2\) and \(P_{{\rm circ},s}={\cal B}_{x,s}e^{2r}P_0/2\).
Here $\phi_{x,s}$ is the carrier round-trip detuning.
Finesse alone does not fix the circulating power: the input coupling
and other losses are needed. At critical coupling and high finesse,
$\mathcal B_x\simeq\mathcal F_x/\pi$.
As an experimental scale reference, LIDA injected $12\,\mathrm W$
into a carrier mode with finesse $7.42\times10^4$ and reported
$118\,\mathrm{kW}$ average circulating power; its signal-mode finesse
was $2.22\times10^3$ \cite{LIDA2024}.

In the bright-seed continuous-wave limit $N_{\rm int}\simeq e^{2r}N_0$;
the exact single-mode count is Eq.~\eqref{eq:interaction_photons_exact}.
Here $N_{\rm int}$
counts photons delivered to the interaction input during the observation,
before repeated cavity passages. At fixed cavity transfer,
the signal-to-noise ratio becomes
\(\mathrm{SNR}_{\rm cl} =\frac{2{\cal C}_m\sqrt{\eta_{\rm ro}N_{\rm int}} \,\psi_{a,{\rm rms}}}{\sqrt{V_y}}\).
For $V_y=1$, this has coherent-state shot-noise scaling. Fixing both
$N_{\rm int}$ and the cavity also fixes the circulating resource.

\subsection{Equal-resource comparison and information bound}
\label{subsec:resource_bound}

First compare exactly equal interaction photon numbers in the
single-mode model. A coherent probe with $|\beta|^2=n_S$, followed
by an $x$-polarization attenuator with $t_x^2=n_0/n_S$ and $t_y=1$,
has the same zero-signal output carrier. For the amplified seed
$n_S>n_0$, so this is a passive attenuation. Its output is coherent,
with mean total photon number
\begin{equation}
 N_{{\rm tot,att}}=n_0+(n_S-n_0)\sin^2\psi_a.
 \label{eq:total_photons_passive}
\end{equation}
Consequently the exact difference between the two detected totals is
$N_{{\rm tot,out}}-N_{{\rm tot,att}}=\nu^2\sin^2\psi_a$.
For $n_0\gg\nu^2$, both totals contain the same dominant term
$n_0(e^{2r}-1)\sin^2\psi_a$. In the passive device this variation
arises because the rotated photons avoid attenuation; including
the rejected port restores its photon-number balance. Thus a
rotation-dependent detected total does not uniquely identify an OPO
advantage. At equal coherent amplitudes $\beta=e^r\alpha$ instead,
the difference is $2\nu^2\sin^2\psi_a$, but that comparison omits
the $\nu^2$ interaction photons of the squeezed probe.

The rotation generator makes the fundamental sensitivity bound explicit. Writing
$U_a(\psi_a)=\exp(-i\psi_a\hat G)$ with the two-mode generator
$\hat G=i(\hat a_y^\dagger\hat a_x-\hat a_x^\dagger\hat a_y)$,
any pure input state of the form $|\chi\rangle_x|0\rangle_y$ yields $\langle\hat G\rangle=0$ and
$\langle\hat G^2\rangle=\langle\hat N_x\rangle$, regardless of how the bright mode $|\chi\rangle_x$ is squeezed. By the quantum Cram\'er--Rao bound \cite{BraunsteinCaves1994}, the classical Fisher information $F_C$ of an arbitrary measurement is bounded by the quantum Fisher information $F_Q$,
\begin{equation}
	F_Q(\psi_a)=4\operatorname{Var}(\hat G)=4n_S,
	\qquad F_C\leq F_Q,
	\label{eq:rotation_qfi_bound}
\end{equation}
where the structure of the encoding generator $\hat G$ is essential \cite{Takeoka2017}. Because the subsequent inverse OPO is independent of $\psi_a$, it acts as a unitary post-processing that cannot increase $F_Q(\psi_a)$, leaving the bound identical to that of a coherent state with the same interaction photon number $n_S$. Note that Eq.~\eqref{eq:rotation_qfi_bound} constrains the single-pass rotation parameter $\psi_a$ of Sec.~\ref{sec:su11_direct_readout} (the common cavity transfer must be folded in when estimating $g_{a\gamma}$), and does not preclude quantum enhancements from squeezing the $y$ signal-input vacuum or in standard single-mode phase sensing.

In the bright-seed, small-modulation limit the OPO and attenuator
versions of the downstream-EOM polarimetric readout therefore obey
\begin{equation}
 \mathrm{SNR}_{\rm OPO}=\mathrm{SNR}_{\rm att}
 =2{\cal C}_m\sqrt{\eta_{\rm ro}N_{\rm int}}\,
 \psi_{a,{\rm rms}}.
 \label{eq:active_passive_equal_resource}
\end{equation}
Carrier deamplification can reduce detector loading, but a selective
attenuator produces the same reference-power reduction. Noise already
converted into $y$ is not removed by an inverse operation on $x$.

Let $\eta_{y,j}$ be the extra signal power transmission and $W_j(f)$
the detected noise PSD normalized to the same reference shot noise.
At fixed circulating and reference powers, with otherwise identical
transfer functions, the narrow-band sensitivities satisfy
\begin{equation}
 \frac{g_{\min,\rm OPO}}{g_{\min,\rm att}}
 =\left[\frac{W_{\rm OPO}\eta_{y,\rm att}}
 {W_{\rm att}\eta_{y,\rm OPO}}\right]^{1/2}.
 \label{eq:resource_noise_ratio}
\end{equation}
For $\eta_{y,\rm OPO}=0.95$ and $\eta_{y,\rm att}=0.99$,
equal noise gives a factor $1.02084$ against the OPO readout.
An improvement requires $W_{\rm OPO}/W_{\rm att}<0.9596$.
We do not assign a favorable technical spectrum to enforce this
inequality. The calculations below instead give the full optical
comparison and the spectral requirements that any implementation
must meet. Frequency-dependent advantages require the same
line-integrated information comparison as the signal itself.

\section{Four-port sensitivity and statistical convention}
\label{sec:numerical-analysis}

We use $L_{\rm rt}=10\,\mathrm m$, $\ell=4.95\,\mathrm m$,
$d=0.05\,\mathrm m$, laser wavelength $\lambda=1064\,\mathrm{nm}$,
$T_{\rm int}=1\,\mathrm{yr}$, $P_0=1\,\mathrm W$,
$r_1=r_2=0.35$ for the first and final OPOs, $\eta_{\rm ro}=0.8$,
${\cal F}_x=7.5\times10^4$, ${\cal F}_y=10^4$ and
$\rho_{\rm local}=0.4\,\mathrm{GeV\,cm^{-3}}$
\cite{deSalasWidmark2021}.
The local axion density $\rho_a$ used below is set equal to
$\rho_{\rm local}$ for these projections.
We denote the signal-mode detuning by $\Delta_y$ and its
amplitude half-linewidth by $\gamma_y$.
All sensitivity curves are evaluated using the four-port transfer
$H_{4p}$ of Appendix~\ref{app:input_output}, including the direct
path to the extraction port. For comparison, the single-port
critical-coupling approximation of Sec.~\ref{sec:ring_cavity}
gives an idealized analytical reference value of
$7.8\times10^{-14}\,\mathrm{GeV}^{-1}$ before accounting for
four-port extraction losses.

Let $\alpha_s$ be the seed photon-flux amplitude in circulation
channel $s$, normalized by $\sum_s|\alpha_s|^2=P_0/(\hbar\omega_0)$,
and let $B_s$ be its residual carrier amplitude after the final OPO.
For the symmetric channels define $b_x=B_s/\alpha_s$ and
$\Phi_B=\sum_s|B_s|^2=P_0|b_x|^2/(\hbar\omega_0)$.
The dimensionless transfer $H_{4p}$ is normalized by
$Y_s(\Omega)=(g_{a\gamma}a_\Omega/2)e^{r_1}\alpha_s
H_{4p}(\Omega)$ before extra signal-path loss.
We calibrate the optimally combined records to the measured output
polarization quadrature $q$, namely the signal field divided by
the residual carrier. For extra signal transmission $\eta_y$,
\begin{equation}
 {\cal K}(f;m_a)=
 \frac{e^{r_1}\sqrt{\eta_y}}{|b_x|}
 \frac{\sqrt{2\rho_{\rm local}}}{2m_a}\,
 H_{\theta,4p}(2\pi f),
 \label{eq:apparatus_coupling_transfer}
\end{equation}
so that a coherent axion oscillation produces peak quadrature
$g_{a\gamma}|{\cal K}|$. Here $H_{\theta,4p}(\Omega)=
[e^{-i\theta}H_{4p}(\Omega)+e^{i\theta}H_{4p}^*(-\Omega)]/2$
is the real-quadrature transfer at the selected reference phase
$\theta$, retaining both optical sidebands as in
Eq.~\eqref{eq:sideband_quadrature}.
A fixed retarder selects the low-frequency signal quadrature;
its angle is kept fixed over the plotted band. The static residual
carrier phase is calibrated in both readouts. Numerical energy units
are converted consistently before reporting couplings in
$\mathrm{GeV}^{-1}$.

In the bright-reference limit the one-sided shot-noise spectrum is
\begin{equation}
 S_{q,\rm shot}=\frac{1}{2\eta_{\rm ro}\Phi_B F_m^{(P)}}.
 \label{eq:output_shot_psd}
\end{equation}
Here $F_m^{(P)}$ is the dimensionless count-likelihood modulation
factor of Eq.~\eqref{eq:coherent_leakage_modulation}, including the
stated leakage backgrounds. It is normalized to unity for ideal
zero-background Poisson weighting and reduces to
${\cal C}_{m,{\rm opt}}^2$ in the corresponding small-angle limit;
it is not assigned unity for a cosine lock-in. The exact count model
and the bound on neglected bright-quadrature corrections are given in
Appendix~\ref{app:input_output}. Additional technical noise is added
in this same output calibration. In the sensitivity formulas, $S_q$
denotes the total one-sided noise PSD in this calibration, equal to
$S_{q,\rm shot}$ in the ideal shot-noise-only case.

For a deterministic sinusoid with many cycles in one coherent
observation, the amplitude convention gives
$g_{\rm coh}^{\rm NE}=\sqrt{S_q/T_{\rm int}}/|{\cal K}|$.
The ideal four-port value is
$1.60\times10^{-13}\,\mathrm{GeV}^{-1}$.
The conditions for its plateau are
$\Omega_aT_{\rm rt}\ll1$, $\Omega_a\tau_{{\rm cav},y}\ll1$,
$T_{\rm int}\lesssim\tau_a$ and $\Omega_aT_{\rm int}\gg1$.
Here $\tau_{{\rm cav},y}=1.062\times10^{-4}\,\mathrm s$,
$\hbar/\tau_{{\rm cav},y}=6.20\times10^{-12}\,\mathrm{eV}$,
and $\tau_a\simeq2\pi\hbar/(m_av_0^2)$ with $v_0=10^{-3}$.
The conventional one-year coherence boundary is
$m_{\rm coh}=1.31\times10^{-16}\,\mathrm{eV}$, or $0.0317\,\mathrm{Hz}$;
many cycles require $m_a\gg1.31\times10^{-22}\,\mathrm{eV}$.
This conventional coherence scale is not a sharp edge. The Maxwell
covariance below gives a smooth crossover; its strictly flat
Gaussian-power limit requires $2\pi\beta_a T_{\rm int}\ll1$,
where $\beta_a=f_0v_0^2/2$ is the axion linewidth parameter defined below.

\subsection{A resolved axion line and finite observation time}
\label{subsec:resolved_line}

For the spectral calculation we use an isotropic, untruncated
Maxwell velocity distribution with $v_0=10^{-3}c$, without a laboratory
boost or annual modulation. Write $f_0=m_a/(2\pi\hbar)$,
$\beta_a=f_0v_0^2/2$ and $x=(f-f_0)/\beta_a$; here $v_0$ is expressed
in units of $c$. The normalized positive-frequency line is
\begin{equation}
 p_m(f)=\frac{2\sqrt{x}\,e^{-x}}{\sqrt\pi\,\beta_a}
 \quad(f\geq f_0),\qquad \int p_m(f)\,df=1.
 \label{eq:maxwell_frequency_line}
\end{equation}
Other halo spectra can be substituted in the same calculation.
With $A=g_{a\gamma}^2$, the signal PSD is
$S_{q,a}=A|{\cal K}|^2p_m/2$. The factor $1/2$ converts the
coherent peak amplitude to its time-averaged power.
For a weak Gaussian signal and $T_{\rm int}$ long compared with
the correlation time, the information is
\begin{equation}
 {\cal I}_{A,\infty}=
 \frac{T_{\rm int}}4\int_0^\infty
 \frac{|{\cal K}(f;m_a)|^4p_m(f)^2}{S_q(f)^2}\,df.
 \label{eq:stochastic_information}
\end{equation}
The displayed noise-equivalent coupling is
$g_{\rm pow}^{\rm NE}={\cal I}_A^{-1/4}$, corresponding to
$A/\sigma_A=1$, where $\sigma_A={\cal I}_A^{-1/2}$ is the
Fisher standard error on $A$. This is not a confidence-level exclusion limit.

\begin{figure*}[t]
 \centering
 \includegraphics[width=\textwidth]{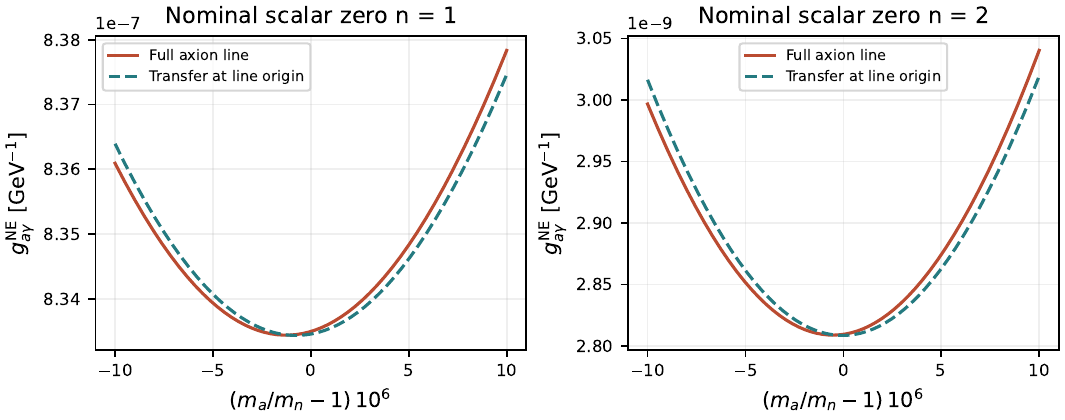}
 \caption{Resolved neighborhoods of the first two nominal zeros
 $m_n=n\,1.23984\times10^{-7}\,\mathrm{eV}$ of the scalar limiting
 kernel. Both curves use the actual four-port optical response and
 the same stochastic normalization. The dashed curve evaluates the
 transfer at the rest-frequency origin of the line; the solid curve
 retains it inside the spectral integral. In this example the ports
 and finite geometry already lift the nominal zeros; the linewidth
 changes the local values and shapes rather than filling an exact
 apparatus zero. The calculation gives finite values
 $8.33\times10^{-7}$ and $2.81\times10^{-9}\,\mathrm{GeV}^{-1}$
 at $m_1$ and $m_2$, respectively.}
 \label{fig:four_port_linewidth}
\end{figure*}

For white calibrated noise and a transfer effectively constant across
the line, the finite-time covariance calculation gives
\begin{equation}
 {\cal I}_A=
 \frac{|{\cal K}|^4T_{\rm int}^2}
 {2S_q^2\left[\sqrt{1+(2\pi\beta_a T_{\rm int})^2}+1\right]}.
 \label{eq:finite_time_power_information}
\end{equation}
Appendix~\ref{app:input_output} specifies how this result is matched
to the full frequency integral. At the coherence transition the
optical response is flat across the axion line; at cavity features
the long-time integral applies with very high accuracy. Since the full four-port transfer is retained inside the spectral integral,
the sensitivity near nominal scalar zeros follows directly from
the apparatus response (Fig.~\ref{fig:four_port_linewidth}).

\begin{figure*}[tp]
 \centering
 \includegraphics[width=\textwidth]{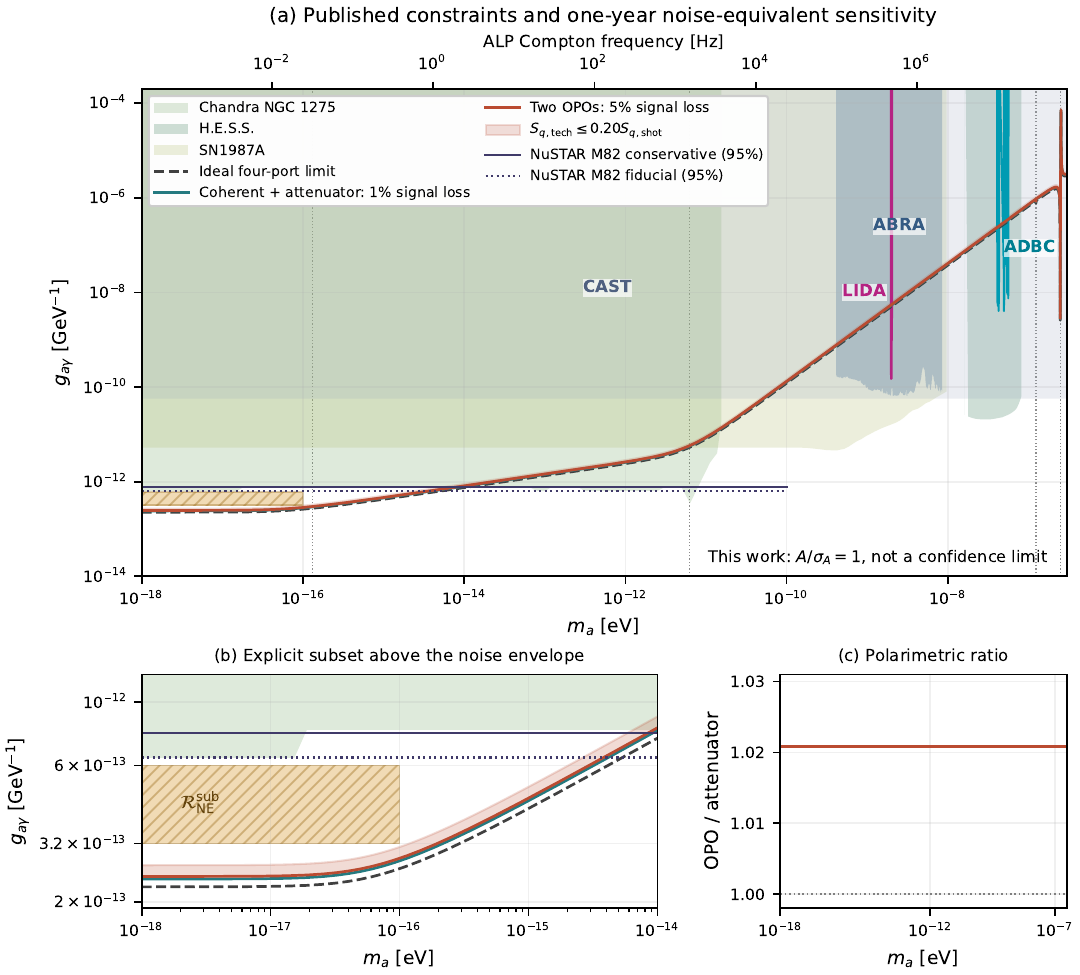}
 \caption{(a) Published bounds and one-year noise-equivalent
polarimetric sensitivity of the four-port apparatus. Existing contours
are CAST, ABRACADABRA-10 cm, LIDA, ADBC, Chandra NGC 1275,
H.E.S.S. and SN1987A
\cite{CAST2024,ABRA2021,LIDA2024,PandeyADBC2024,Chandra2020,HESS2013,SN1987A2015}.
The NuSTAR M82 horizontal segments show the approximate low-mass
95\% confidence limits, $7.8\times10^{-13}\,\mathrm{GeV}^{-1}$
(conservative) and $6.4\times10^{-13}\,\mathrm{GeV}^{-1}$ (fiducial),
both from Ref.~\cite{NingSafdi2025} and shown only up to
$m_a\lesssim10^{-10}\,\mathrm{eV}$; they do not represent the
high-mass contour. Source, halo and confidence assumptions
remain those of each cited analysis. Our curves use $A/\sigma_A=1$,
the Maxwell line and $\rho_a=0.4\,\mathrm{GeV\,cm^{-3}}$.
The two solid curves match circulating and residual-reference powers,
with splitting $\Delta_y/\gamma_y=0.1$, analyzer leakage $\epsilon=10^{-6}$,
coherent leakage $\xi=10^{-4}$, and extra signal transmissions $0.99$
(attenuator) and $0.95$ (OPOs). The dashed curve is the ideal four-port
limit. The band allows $S_{q,\rm tech}\leq0.20S_{q,\rm shot}$.
(b) Low-mass detail and the explicit subset in Eq.~\eqref{eq:prospective_subset}, hatched; this defines a noise-equivalent prospective sensitivity region, rather than a confidence contour. (c) The polarimetric coupling ratio is
$1.02084$. Vertical markers in (a) indicate the conventional coherence
boundary, signal-cavity bandwidth and first two nominal scalar zeros.}

 \label{fig:exclusion_plot}
\end{figure*}

In the coherent limit $g_{\rm pow}^{\rm NE}=\sqrt2\,g_{\rm coh}^{\rm NE}$.
This numerical factor comes from changing from a sinusoidal-amplitude
criterion to a Gaussian-power criterion. Accordingly, the ideal
four-port plateau in Fig.~\ref{fig:exclusion_plot} is
$2.26\times10^{-13}\,\mathrm{GeV}^{-1}$. With the imperfections
specified in Sec.~\ref{subsec:spectral_requirements}, it is
$2.45\times10^{-13}\,\mathrm{GeV}^{-1}$; allowing the stated
technical-noise budget raises it to at most
$2.69\times10^{-13}\,\mathrm{GeV}^{-1}$.
The amplitude-convention value for that same imperfect example is
$1.74\times10^{-13}\,\mathrm{GeV}^{-1}$.

Historical solar-axion helioscope constraints include the
CAST $^3$He data \cite{Axion10}.
The NuSTAR M82 analysis of Ning and Safdi \cite{NingSafdi2025}
reports low-mass 95\% confidence upper limits of
$7.8\times10^{-13}\,\mathrm{GeV}^{-1}$ (conservative) and
$6.4\times10^{-13}\,\mathrm{GeV}^{-1}$ (fiducial), both for
$m_a\lesssim10^{-10}\,\mathrm{eV}$. Both limits appear in the
published analysis and are shown in Fig.~\ref{fig:exclusion_plot};
the conservative limit is used in the comparison below. These limits rely on stellar axion production
and magnetic-field conversion models \cite{Raffelt1996,Raffelt2008},
rather than on the local dark-matter density assumed for our optical signal.

To specify the prospective coverage, let $B_{\rm disp}(m_a)$ denote
the lower envelope of the displayed bounds, with each bound restricted
to its stated mass range. Let $g_{\rm env}^{\rm NE}(m_a)$ be the
two-OPO polarimetric curve including
$S_{q,\rm tech}=0.20S_{q,\rm shot}$. We define
\begin{equation}
 \mathcal R_{\rm NE}=\{(m_a,|g_{a\gamma}|):
 g_{\rm env}^{\rm NE}(m_a)\leq|g_{a\gamma}|<B_{\rm disp}(m_a)\}.
 \label{eq:prospective_region}
\end{equation}
An explicit, deliberately nonmaximal window within this prospective
sensitivity region is
\begin{align}
 10^{-18}\leq &\,m_a/\mathrm{eV}\leq10^{-16},\nonumber\\
 3.2\times10^{-13}\leq &\,|g_{a\gamma}|/(\mathrm{GeV}^{-1})
                         <6.0\times10^{-13}.
 \label{eq:prospective_subset}
\end{align}
Across this mass interval the noise envelope rises from
$2.69$ to $3.11\times10^{-13}\,\mathrm{GeV}^{-1}$; the upper edge
of the subset remains below all the bounds shown, including both
quoted M82 limits. The hatched area in
Fig.~\ref{fig:exclusion_plot}(b) therefore defines a noise-equivalent
prospective sensitivity region; it is not a 95\% detection or exclusion
forecast, and makes no assertion about constraints absent from the
figure. A confidence construction would require the signal likelihood,
the nonnegative-power boundary and nuisance parameters; a universal
rescaling of the Fisher threshold is not justified in the unresolved-line
regime. The optical reach assumes
$\rho_a=0.4\,\mathrm{GeV\,cm^{-3}}$; for another local axion density,
its lower boundary scales as $\rho_a^{-1/2}$.

The displayed band reaches $f_0=72.5\,\mathrm{MHz}$ and assumes
readout of the modulation sidebands with sufficient temporal bandwidth
for the stated likelihood weighting. Detector roll-off, electronic
noise and imperfect implementation of that weighting must enter
$S_q(f)$ and the calibrated signal transfer. They have not been
assigned favorable measured values. A real instrument's validated
readout bandwidth may restrict the usable part of the curve.

\section{Noise budget and parameterized input--output study}
\label{sec:noise_equivalent_coupling}

The quantitative comparison uses the calibrated output quadrature $q$
of Sec.~\ref{sec:numerical-analysis}. Its one-sided PSD is
\begin{equation}
 S_q^{\rm tot}(f)=S_{q,\rm shot}(f)+S_{q,\rm tech}(f),
 \label{eq:noise_psd_sum}
\end{equation}
in $\mathrm{rad}^2/\mathrm{Hz}$. Mirror, polarization, modulation,
leakage and control disturbances must be propagated into this same
calibration. Channel correlations require the noise matrix $C_n$
of Sec.~\ref{sec:ring_cavity}. We first isolate static optical
penalties and then give spectral requirements for the main curve.

\subsection{Mirror birefringence and quadrature selection}
\label{subsec:mirror_birefringence_realistic}

Mirror birefringence is generally a retarder rather than a scalar
rotation. For eigenaxes at angle $\beta_{\rm mir}$ and differential phase
$\Delta\varphi_{\rm mir}$, let $R(\beta)$ denote the ordinary
two-dimensional rotation matrix, distinct from the reflection matrix
$R$ of Sec.~\ref{sec:ring_cavity}. An effective element is
\begin{equation}
 J_{\rm mir}=R(-\beta_{\rm mir})
 \begin{pmatrix}
 e^{i\Delta\varphi_{\rm mir}/2}&0\\
 0&e^{-i\Delta\varphi_{\rm mir}/2}
 \end{pmatrix}R(\beta_{\rm mir}).
 \label{eq:mirror_jones_retarder}
\end{equation}
A carrier aligned with an eigenaxis acquires no orthogonal component
from this retardance alone. Leakage depends on axis misalignment,
geometric transport and unequal losses. The retardance also splits
the two cavity resonances, changing both signal amplitude and phase.

DANCE, LIDA and related control studies address these effects using
auxiliary resonance control, heterodyne treatment of splitting,
wavelength tuning or reflective phase elements
\cite{FujimotoDANCEAuxiliary,LIDA2024,Takidera2025,Kuramoto2025,APE2026}.
For the present readout, the signal quadrature is determined by both
optical sidebands and the EOM reference. A real EOM rotation does not
automatically measure a quadrature-phase signal. The convention and
sideband projection are specified in Appendix~\ref{app:input_output}.
External modulation translates mirror noise as well as signal;
its rejection must follow from the measured channel responses.

\subsection{Explicit four-port example}
\label{subsec:four_port_study}

To make the input--output comparison quantitative, choose
$\ell=4.95\,\mathrm m$, $d=0.05\,\mathrm m$, four identical mirrors
without absorption, and the parameters in
Sec.~\ref{sec:numerical-analysis}. For polarization $j$ each
mirror has power transmission $T_j=1-\sqrt{q_j}$, giving
$T_x=20.944$ ppm and $T_y=157.067$ ppm. The vacuum fields of all
four ports are retained. These are specified model parameters,
not measured coatings. The exact cyclic propagation and covariance
equations are given in Appendix~\ref{app:input_output}.

The output carrier amplitude transfer is $h_x=0.500005$.
The model gives $12.019\,\mathrm{kW}$ circulating power per direction,
$0.250005\,\mathrm W$ total residual carrier after the inverse OPOs,
and a mean detected modulation-reference power of
$10.000\,\mu\mathrm W$ for $\psi_0=10^{-2}$.
The effective low-mass length is $4.27166\times10^4\,\mathrm m$.
With the signal-preserving sum and optimal weighting,
\(g_{a\gamma,4p}^{\rm NE}\simeq 1.60\times10^{-13}\,\mathrm{GeV}^{-1}\)
in the deterministic coherent-amplitude convention. Its difference from
the ideal normalization quoted in Sec.~\ref{sec:numerical-analysis} follows mainly from allocating
transmission to four ports, rather than the critical-coupling
normalization. It is a conditional example, not a cavity-independent sensitivity
at fixed finesse.

Table~\ref{tab:parameter_penalties} varies post-cavity signal loss
$\ell_y$, polarization splitting $\delta=\Delta_y/\gamma_y$, residual
second-pump phase $\phi$, and analyzer power leakage $\epsilon$.
Here $\gamma_y/(2\pi)=1498.96\,\mathrm{Hz}$; $\delta=0.1,0.3,1$
corresponds to splittings of approximately $150,450,1499\,\mathrm{Hz}$.
The reference is the same four-port system with ideal leading-order
small-modulation readout. Static splitting is compensated by selecting
the measured quadrature; the residual pump-phase error is not
compensated. This static table sets the coherent parasitic leakage $\xi=0$; the main broad-band
comparison below instead includes coherent leakage and calibrates the
static carrier phase in both instruments.

\begin{table}[htbp]
 \centering
 \caption{Quasistatic sensitivity penalties for the four-port model.
 The factor multiplies the ideal four-port reference sensitivity
 $g_{\min,0}$ quoted in the text before rounding.
 The combined row uses $(\ell_y,|\delta|,\phi,\epsilon)
 =(0.05,0.1,0.01,10^{-6})$, with $\phi$ in radians.}
 \label{tab:parameter_penalties}
 \small
 \begin{tabular}{@{}lc@{}}
 \toprule
 Variation & $g_{\min}/g_{\min,0}$\\
 \midrule
 Signal loss $\ell_y=0.05$ & 1.0260\\
 Splitting $|\delta|=0.1$ & 1.0050\\
 Splitting $|\delta|=0.3$ & 1.0440\\
 Splitting $|\delta|=1$ & 1.4141\\
 Pump phase $\phi=0.1\,\mathrm{rad}$ & 1.0013\\
 Power leakage $\epsilon=10^{-6}$ & 1.0538\\
 Power leakage $\epsilon=10^{-5}$ & 1.1966\\
 Power leakage $\epsilon=10^{-4}$ & 1.8479\\
 Combined & 1.0866\\
 \bottomrule
 \end{tabular}
\end{table}

\begin{figure*}[t]
	\centering
	\includegraphics[width=\textwidth]{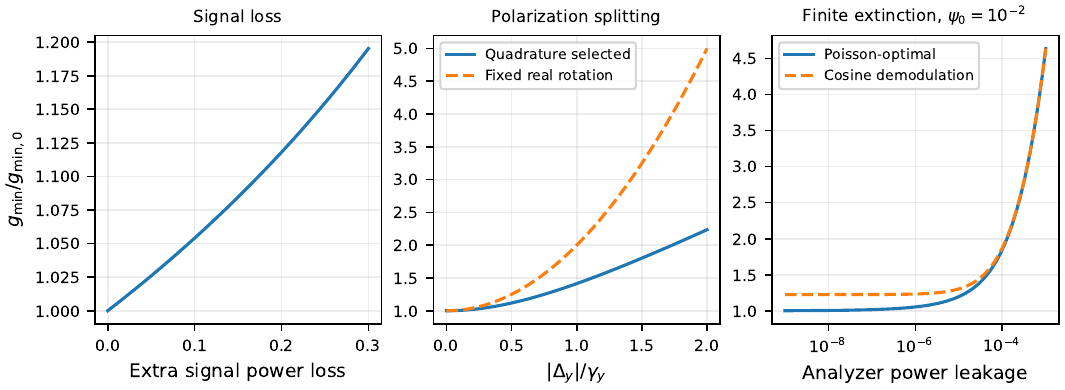}
	\caption{Sensitivity penalties relative to ideal readout: additional
		signal-path loss, quasistatic polarization splitting, and finite
		analyzer power extinction at $\psi_0=10^{-2}$. The central panel
		distinguishes selected quadrature from an uncompensated real
		rotation. The right panel distinguishes Poisson-optimal weighting
		from ordinary cosine demodulation. These curves isolate the stated
		effects and do not include empirical technical-noise spectra.}
	\label{fig:parameter_penalties}
\end{figure*}

With splitting and optimized quadrature the leading penalty is
$\sqrt{1+\delta^2}$; retaining only an uncompensated real-rotation
quadrature instead gives $1+\delta^2$ (see Fig.~\ref{fig:parameter_penalties} for the continuous dependence across loss, splitting, and analyzer extinction). The combined case gives
$G_{\rm sig}=1.38314$, $V_y^{\rm pre}=1$,
$\overline P_{\rm ref,det}=10.0007\,\mu\mathrm W$, and
$g_{a\gamma}^{\rm NE}\simeq1.74\times10^{-13}\,\mathrm{GeV}^{-1}$.
Here $\epsilon$ is orthogonally polarized power leakage collected by
the detector; coherent leakage in the signal polarization requires a
separate complex amplitude.

\subsection{Spectral stability requirements}
\label{subsec:spectral_requirements}

We supplement the static study with quantitative requirements on
technical spectra. For the broad-band comparison choose
$\eta_{y,\rm OPO}=0.95$, $\eta_{y,\rm att}=0.99$,
$\Delta_y/\gamma_y=0.1$, $\epsilon=10^{-6}$ and a coherent
quadrature parasitic field $E_{\rm par}/B=i\xi$ with $\xi=10^{-4}$.
The last field is distinct from orthogonally polarized analyzer
leakage, and its photon background is included in the likelihood.
The OPO phase set point is $\phi_0=0.01$ rad; its static carrier
phase is calibrated, while its fluctuations remain noise inputs.
The passive comparator is adjusted to the same carrier magnitude
and measured quadrature. Both configurations have
$12.019\,\mathrm{kW}$ circulating per direction and
$10.0007\,\mu\mathrm W$ mean detected modulation-reference power.
The broad-band leading count model gives $F_m^{(P)}=0.899955$ and
\begin{equation}
 \sqrt{S_{q,\rm shot}}=7.20\times10^{-10}\,
 \mathrm{rad}/\sqrt{\mathrm{Hz}}.
 \label{eq:required_shot_asd}
\end{equation}

Let $\delta\phi_p$ be residual inverse-pump phase noise. A parasitic
field whose phase does not follow that carrier produces
$\delta q\simeq-\kappa_p\xi\,\delta\phi_p$, where
$\kappa_p=(e^{2r}-1)/2=0.50688$ at small phase error,
with $r=r_2=0.35$ the final-OPO squeezing parameter.
If the parasitic field follows the carrier exactly, this particular
coupling is absent. Let $\kappa_I$ be the separately calibrated
coupling from residual relative-intensity fluctuations to $q$;
perfect intensity normalization would set it to zero. For
uncorrelated mechanisms a concrete budget is
\begin{align}
 S_{q,\rm tech}={}&S_{q,\rm pol}
 +(\kappa_p\xi)^2S_{\phi_p}
 +\kappa_I^2S_{\rm RIN}+S_{q,\rm ro},
 \label{eq:technical_psd_requirements}
\end{align}
where $S_{q,\rm pol}$ is the calibrated polarization-noise PSD,
$S_{\phi_p}$ the pump-phase-noise PSD, $S_{\rm RIN}$ the
relative-intensity-noise PSD, and $S_{q,\rm ro}$ the residual
readout-noise PSD, including EOM and electronic disturbances.
Correlations require the corresponding
cross-spectral terms; common polarization noise is not canceled by
the signal-preserving sum.

For at most a factor $1.10$ sensitivity degradation it is sufficient
that $S_{q,\rm tech}(f)\leq0.21S_{q,\rm shot}(f)$ throughout each
analyzed axion line. We allocate $0.05S_{q,\rm shot}$ to each of the
four terms, giving a total $0.20$ and a factor $\sqrt{1.20}=1.0954$.
The resulting requirements are listed in Table~\ref{tab:spectral_requirements}.
\begin{table}[htbp]
 \centering
 \caption{One-sided amplitude-spectral-density requirements, applied
 at every analyzed frequency. They are tolerances, not measured PSDs.
 The pump and intensity rows use $\xi=10^{-4}$ and
 $\kappa_I=10^{-4}$ rad, respectively.}
 \label{tab:spectral_requirements}
 \small
 \begin{tabular}{@{}ll@{}}
 \toprule
 Residual disturbance & Required ASD\\
 \midrule
 Output polarization & $1.61\times10^{-10}\,\mathrm{rad}/\sqrt{\mathrm{Hz}}$\\
 Pump phase & $3.18\times10^{-6}\,\mathrm{rad}/\sqrt{\mathrm{Hz}}$\\
 Relative intensity & $1.61\times10^{-6}/\sqrt{\mathrm{Hz}}$\\
 Calibrated readout & $1.61\times10^{-10}\,\mathrm{rad}/\sqrt{\mathrm{Hz}}$\\
 \bottomrule
 \end{tabular}
\end{table}
The pump tolerance scales as $10^{-4}/|\xi|$ and the intensity
tolerance as $10^{-4}\,\mathrm{rad}/|\kappa_I|$. Values outside
these illustrative couplings require recalculating both the background
and the transfer. The window in Eq.~\eqref{eq:prospective_subset}
corresponds to $f_0\simeq0.24$--$24\,\mathrm{mHz}$. LIDA provides
an experimental reference for the circulating-power scale discussed in
Sec.~\ref{sec:power_accounting}, whereas its published search near
$480\,\mathrm{kHz}$ does not establish the polarization stability
required in this much lower frequency band \cite{LIDA2024}.
The requirements in Table~\ref{tab:spectral_requirements} must
therefore be verified in the axion analysis band using the calibrated
cavity and demodulation response. External modulation does not remove
a polarization disturbance that has the same calibrated response
as the axion.

This budget supplies testable conditions for approaching the optical
curve. It does not establish an OPO noise advantage: the passive
reference must be characterized in the same calibration and compared
using Eq.~\eqref{eq:resource_noise_ratio} or the full spectral integral.

\section{Comparison with optical axion searches}
\label{sec:comparison_realistic}

DANCE, LIDA and ADBC establish cavity-enhanced axion polarimetry with
different geometries and polarization-mode controls
\cite{Obata2018,DANCEAct1,LIDA2024,LiuADBC2019,PandeyADBC2024}.
Decade-long blazar polarimetry probes ultralight axion dark matter at
smaller masses \cite{Huang2026}, outside the displayed comparison range.
Mirror transport, birefringence and imperfections have been studied
in this context \cite{GarciaCely2025,Takidera2025,Kuramoto2025,APE2026}.
We compare the squeeze--rotate--unsqueeze sequence with passive and
coherent readouts, using the total-photon observable, the extracted-field
transfer and the associated noise spectra. The quantum-information bound
is evaluated for the polarization-rotation encoding considered here.

Unlike the scalar-field single-arm proposal
\cite{CapolupoScalar2026}, the signal here is an off-resonant rotation
between two polarizations. Unlike the signal-vacuum squeezing of
Martynov and Miao \cite{MartynovMiao2020}, the bright-mode OPOs leave
the signal vacuum unsqueezed at the operating null.
Figure~\ref{fig:exclusion_plot} includes the NuSTAR M82 limits and the
prospective sensitivity region of Eq.~\eqref{eq:prospective_subset}; the common-resource
comparison is treated separately from these external bounds.

\subsection{Photon counting and continuous-variable polarimetry}
\label{sec:photon_counting_lockin}

The photon-counting proposal of Yu et al. \cite{Yu2024} suppresses
the bright polarization and measures photons in the orthogonal mode,
with $N_{\rm sig}\propto g_{a\gamma}^2$. Sufficient extinction can
bring the background toward the detector dark-count level. Here the
EOM reference instead yields a linear interference signal at the cost
of reference shot noise and modulation stability. These readouts trade
different backgrounds against different signal observables.

The projections in Ref.~\cite{Yu2024} span cavity lengths from
metres to kilometres. In the quoted $0.1$--$1\,\mathrm{neV}$ range,
the tuning relation $f_0=c/(2L)$ corresponds to about
$6.2\,\mathrm{km}$--$620\,\mathrm m$, with length-dependent
circulating-power assumptions reaching $10\,\mathrm{MW}$.
They should therefore not be interpreted as projections for a
fixed $10\,\mathrm m$ cavity. Table~\ref{tab:comparison} provides
context; the controlled optical-resource comparison is
Fig.~\ref{fig:tc_sensitivity}.

\begin{table*}[t]
 \centering
 \caption{Indicative comparison of searches and projections with
 different source populations, observables and confidence conventions.
 The two entries for this work refer to the same four-port example
 with deterministic coherent-amplitude or Gaussian stochastic-power
 information, respectively. They are noise-equivalent sensitivities
 with optimal weighting, not exclusion bounds. The coherent plateaus also require the conditions stated in
 Sec.~\ref{sec:numerical-analysis}.}
 \label{tab:comparison}
 \scriptsize
 \renewcommand{\arraystretch}{1.18}
 \setlength{\tabcolsep}{3pt}
 \begin{tabular*}{\textwidth}{@{\extracolsep{\fill}}llll@{}}
 \toprule
 Experiment/proposal & Technique/scale & Mass range & Coupling result/scale\\
 \midrule
 CAST \cite{CAST2024}
 & solar helioscope & $m_a\lesssim0.02\,\mathrm{eV}$
 & $g_{a\gamma}<5.8\times10^{-11}\,\mathrm{GeV}^{-1}$ (95\% C.L.)\\
 NuSTAR M82 (conservative) \cite{NingSafdi2025}
 & \shortstack[l]{stellar production;\\X-ray conversion} & $m_a\lesssim10^{-10}\,\mathrm{eV}$
 & $g_{a\gamma}\lesssim7.8\times10^{-13}\,\mathrm{GeV}^{-1}$ (95\% C.L.)\\
 IAXO \cite{IAXO2019}
 & next-generation helioscope & solar-axion reach
 & $g_{a\gamma}\sim10^{-12}\,\mathrm{GeV}^{-1}$ projected\\
 DANCE Act-1 \cite{DANCEAct1}
 & bow-tie polarimetry & $10^{-14}$--$10^{-13}\,\mathrm{eV}$
 & $\lesssim8\times10^{-4}\,\mathrm{GeV}^{-1}$\\
 LIDA \cite{LIDA2024}
 & laser-interferometric polarimetry & $1.97$--$2.01\,\mathrm{neV}$
 & $1.51\times10^{-10}\,\mathrm{GeV}^{-1}$\\
 ADBC \cite{PandeyADBC2024}
 & tunable birefringent cavity
 & \shortstack[l]{three windows within\\$40.9$--$56.7\,\mathrm{neV}$}
 & $1.9\times10^{-8}\,\mathrm{GeV}^{-1}$\\
 Yu et al. \cite{Yu2024}
 & \shortstack[l]{photon counting; length-dependent\\metre-to-kilometre configurations}
 & $0.1$--$1\,\mathrm{neV}$
 & $10^{-17}$--$10^{-16}\,\mathrm{GeV}^{-1}$ projected\\
 This work: coherent amplitude
 & \shortstack[l]{$10\,\mathrm m$; four-port model;\\$\ell=4.95\,\mathrm m$, $d=0.05\,\mathrm m$}
 & coherent plateau below $m_{\rm coh}$
 & \shortstack[l]{$1.60\times10^{-13}\,\mathrm{GeV}^{-1}$ ideal;\\$1.74\times10^{-13}\,\mathrm{GeV}^{-1}$ with imperfections}\\
 This work: stochastic power
 & \shortstack[l]{same four ports; resolved axion line;\\one year; $A/\sigma_A=1$, $A=g_{a\gamma}^2$}
 & full curve in Fig.~\ref{fig:exclusion_plot}
 & \shortstack[l]{plateau: $(2.26,2.45,2.69)\times10^{-13}\,\mathrm{GeV}^{-1}$;\\ideal, imperfections, technical allowance}\\
 \bottomrule
 \end{tabular*}
\end{table*}

\section{Bias-assisted linear readout of the total photon number}
\label{sec:bias_total_counts}

The total outgoing photon number provides an additional observable for the
squeeze--rotation--antisqueeze sequence. At zero bias its change is even
in the axion-induced rotation and vanishes to first order at the ideal
bright-mode closure point. A calibrated
polarization bias, applied after the interaction and before the inverse
OPO, moves the apparatus away from this stationary point and produces a
linear response in the sum of the two polarization counts. We derive
this response, its realization with a phase shifter, and the noise of
this readout for the extracted cavity fields.

\subsection{Phase-shifter implementation and placement}

We use the real rotation convention
\begin{equation}
 R(\theta)=\begin{pmatrix}\cos\theta&-\sin\theta\\
 \sin\theta&\cos\theta\end{pmatrix}.
 \label{eq:tc_rotation}
\end{equation}
Starting from the unbiased, unmixed output, a phase shifter acting
only as $a_x\mapsto e^{i\varphi}a_x$, with $a_y$ unchanged,
does not provide this bias: the total mean remains
even in the small rotation $\psi_a$. Likewise, any lossless passive
optical transformation placed after the final OPO preserves
$N_x+N_y$ exactly.

The required element instead imposes opposite phases on the circular
polarizations. With
$a_+=(a_x+i a_y)/\sqrt{2}$ and $a_-=(a_x-i a_y)/\sqrt{2}$, choose
\begin{equation}
 a_+\longmapsto e^{i\theta_b}a_+,
 \qquad a_-\longmapsto e^{-i\theta_b}a_- .
 \label{eq:tc_circular_bias}
\end{equation}
This is precisely $R(\theta_b)$ in the linear basis. A differential
phase $\delta_b=2\theta_b$ can therefore be implemented by a calibrated
retarder between polarization-basis converters, or by an equivalent
polarization rotator. A retarder between the linear $x,y$ axes alone
is insufficient. At the operating point used below,
$\theta_b=\pi/4$ and $\delta_b=\pi/2$.
The order of operations is first OPO, cavity interaction and extraction,
quadrature calibration, bias rotation, inverse OPO, and detection.
The existing downstream polarimetric modulation does not change the
total counts and is not the bias considered here.

\subsection{Exact ideal-mode result}

First consider the lossless single-mode sequence with equal OPO
parameters $r$, phase-matched real seed amplitude $\alpha$, and
$n_0=|\alpha|^2$. The first OPO amplifies the mean $x$ field by $e^r$;
the second deamplifies this same quadrature by $e^{-r}$. For a real
interaction rotation, $R(\theta_b)R(\psi_a)=R(\theta_b+\psi_a)$.
The quadrature selector is the identity in this ideal example.
Defining $\nu=\sinh r$ and $u=\theta_b+\psi_a$, the output means are
\begin{align}
 \langle N_x\rangle&=n_0\cos^2u+\nu^2\sin^2u,\\
 \langle N_y\rangle&=(e^{2r}n_0+\nu^2)\sin^2u.
 \label{eq:tc_individual_counts}
\end{align}
Consequently,
\begin{align}
 \langle N_{\rm tot}\rangle&=n_0+\mathcal{A}_N\sin^2u,\\
 \mathcal{A}_N&=n_0(e^{2r}-1)+2\sinh^2r.
 \label{eq:tc_total_ideal}
\end{align}
These expressions include spontaneous photons in the accepted mode.
For a continuous-wave detector their contribution must be integrated
over the accepted spatial and temporal modes; $\sinh^2r$ is not a
photon flux without a specified bandwidth.

Subtracting the background measured at $\psi_a=0$ gives
\begin{align}
 \Delta\langle N_{\rm tot}\rangle
 &=\mathcal{A}_N\sin(2\theta_b)\psi_a
   +\mathcal{A}_N\cos(2\theta_b)\psi_a^2+O(\psi_a^3).
 \label{eq:tc_linear_counts}
\end{align}
Thus the bias point $\theta_b=\pi/4$ maximizes the slope at fixed $n_0,r$ and eliminates
the quadratic term:
\begin{align}
 \langle N_{\rm tot}\rangle_0&=n_0+\mathcal{A}_N/2,\\
 \Delta\langle N_{\rm tot}\rangle
 &=\frac{\mathcal{A}_N}{2}\sin(2\psi_a)
 =\mathcal{A}_N\psi_a-\frac{2\mathcal{A}_N}{3}\psi_a^3+O(\psi_a^5).
 \label{eq:tc_quarter_bias}
\end{align}
Reversing the bias reverses the linear response. The axion rotation
itself conserves photon number; the changed total output is associated
with energy exchange with the parametric pumps. A nonzero slope
therefore reveals the active number transduction, rather than photon
creation by the passive rotation. The maximum-slope point need not
minimize the noise-equivalent rotation.

Evaluating the single-mode variance of Appendix~\ref{app:bright_mode_closure}
[Eq.~\eqref{eq:total_photon_variance}] at the biased angle $u=\theta_b+\psi_a$
gives
\begin{align}
 \operatorname{Var}N_{\rm tot}
 &=n_0\left[(\cos^2u+e^{2r}\sin^2u)^2
       +4\nu^2\cos^2u\sin^2u\right]\nonumber\\
 &\quad+4\nu^2\sin^2u(1+\nu^2\sin^2u).
 \label{eq:tc_ideal_variance}
\end{align}
It is generally not Poissonian. With equal independent detector
quantum efficiency $\eta_d$, thinning gives
$\operatorname{Var}N_d=\eta_d^2\operatorname{Var}N_{\rm tot}
+\eta_d(1-\eta_d)\langle N_{\rm tot}\rangle$.
The cavity output requires its own covariance, considered below;
one cannot simply substitute its carrier power into the lossless
single-mode variance.

\subsection{Coefficient for the actual extracted fields}

Normalize coherent amplitudes to the square root of photon flux.
In circulation channel $s$, the carrier amplitude $C_s$ is made real
and positive by a fixed phase calibration. Write the weak orthogonal
mean envelope as $Y_s(t)$, with Fourier convention
$Y_s(t)=\int Y_s(\Omega)e^{-i\Omega t}d\Omega/(2\pi)$.
For the four-port geometry of Appendix~\ref{app:input_output},
\begin{align}
 C_s&=h_x e^{r_1}\alpha_s,\qquad h_x=\frac{T_x}{1-q_x},
 \label{eq:tc_extracted_carrier}\\
 Y_s(\Omega)&=e^{r_1}\alpha_s\sqrt{\eta_y}\,
 \frac{g_{a\gamma}a_\Omega}{2}H_{4p}(\Omega).
 \label{eq:tc_extracted_fields}
\end{align}
The extra signal power transmission $\eta_y$ is applied before
bias mixing. The exact $H_{4p}$ is given in
Eq.~\eqref{eq:closed_four_port_transfer}; it includes the segment
to the extraction mirror. Neither field in
Eqs.~\eqref{eq:tc_extracted_carrier} and~\eqref{eq:tc_extracted_fields}
is the circulating cavity field.

Include a fixed signal-quadrature retarder by defining
$\widetilde Y_s(t)=e^{-i\chi}Y_s(t)$. The biased mean fields are
\begin{align}
 b_{x,s}(t)&=C_s\cos\theta_b-\widetilde Y_s(t)\sin\theta_b,
 \nonumber\\
 b_{y,s}(t)&=C_s\sin\theta_b+\widetilde Y_s(t)\cos\theta_b.
 \label{eq:tc_biased_means}
\end{align}
For an ideal inverse OPO with pump-phase error $\phi$, the outgoing
mean amplitudes are
$B_{x,s}=\cosh r_2\,b_{x,s}-e^{i\phi}\sinh r_2\,b_{x,s}^{*}$
and $B_{y,s}=b_{y,s}$. Expansion of
$\hbar\omega_0(|B_{x,s}|^2+|B_{y,s}|^2)$ gives the term
linear in the weak signal and proportional to the carrier flux:
\begin{align}
 \delta P_{\Sigma,s}(t)
 &=\hbar\omega_0 C_s\sin(2\theta_b)
    \operatorname{Re}\{D(\phi)\widetilde Y_s(t)\},
 \label{eq:tc_linear_power}\\
 D(\phi)&=1-\cosh(2r_2)+\sinh(2r_2)e^{-i\phi}.
 \label{eq:tc_complex_gain}
\end{align}
This is a bright-field expression. Contributions of spontaneous
photons, retained in the exact single-mode result above, would
require an accepted-bandwidth model in continuous operation.

The real part in Eq.~\eqref{eq:tc_linear_power} acts on the
\emph{time-domain envelope}, not separately on a complex Fourier
coefficient. Since $a(t)$ is real, the corresponding frequency
kernel is
\begin{align}
 {\cal H}_{\phi,\chi}(\Omega)
 &=\tfrac12\big[D(\phi)e^{-i\chi}H_{4p}(\Omega)
 \nonumber\\
 &\hspace{1.9em}+D(\phi)^*e^{i\chi}H_{4p}^*(-\Omega)\big].
 \label{eq:tc_pump_sidebands}
\end{align}
For $\phi=0$, set $d_2=1-e^{-2r_2}>0$, so
${\cal H}_{0,\chi}=d_2H_\chi$ with
\begin{equation}
 H_\chi(\Omega)=\frac{e^{-i\chi}H_{4p}(\Omega)
                 +e^{i\chi}H_{4p}^*(-\Omega)}{2}.
 \label{eq:tc_two_sidebands}
\end{equation}
Thus the transfer between real time records can retain a
frequency-dependent temporal phase; replacing it by $|H_{4p}|$ would generally give an incorrect
quadrature response.

Define $\psi_{{\rm ext},s}(t)=
\operatorname{Re}\widetilde Y_s(t)/C_s$ and
$P_{{\rm ext},s}=\hbar\omega_0C_s^2$. At zero pump-phase error,
\begin{equation}
 \delta P_{\Sigma,s}(t)=P_{{\rm ext},s}d_2\sin(2\theta_b)
                         \psi_{{\rm ext},s}(t).
 \label{eq:tc_power_gain}
\end{equation}
The port transports specified in this paper give equal signal signs
in the two calibrated channels, so their records are summed and
$\psi_{{\rm ext},s}=\psi_{\rm ext}$. A change of extraction geometry
requires recalculating these transports and relative signs.

For $P_0=1\,\mathrm W$ summed over the seeds, $r_1=r_2=0.35$,
and the stated cavity, the two extracted carriers have total power
\begin{equation}
 P_{\rm ext}=|h_x|^2e^{2r_1}P_0=0.503449\,\mathrm W.
\end{equation}
At $\theta_b=\pi/4$, zero pump-phase error and zero splitting,
\begin{align}
 P_{\Sigma,0}&=\tfrac12(1+e^{-2r_2})P_{\rm ext}
                  =0.376727\,\mathrm W,\\
 \delta P_\Sigma&=(0.253443\,\mathrm{W\,rad^{-1}})
                                      \psi_{\rm ext}.
 \label{eq:tc_numerical_gain}
\end{align}
For a sinusoidal axion field of peak amplitude
$\sqrt{2\rho_a}/m_a$, the low-frequency peak normalization is
$|\psi_{\rm ext}|=0.536698\,[g_{a\gamma}/(1\,\mathrm{GeV}^{-1})]$
for $\rho_a=0.4\,\mathrm{GeV\,cm^{-3}}$, unit signal transmission
and $\chi=0$. This limit requires a flat cavity response as well
as $m_aL_{\rm rt}\ll1$.
Thus $g_{a\gamma}=10^{-13}\,\mathrm{GeV}^{-1}$ gives a power
oscillation of amplitude $1.36023\times10^{-14}\,\mathrm W$.
This coefficient is for a static $\pi/4$ bias; it is not the much
smaller coefficient of a small modulation bias.

\subsection{Direct detection and an optional modulated bias}

After the inverse OPO, a polarization splitter sends $x$ and $y$ to
separate photodiodes. The measured observable is their calibrated
\emph{sum}, not their difference. Summing also over circulation
channels, the photon-flux record is
\begin{equation}
 \widehat\Phi_\Sigma(t)=\sum_s\left[
 \frac{i_{x,s}(t)}{e\eta_{x,s}}+
 \frac{i_{y,s}(t)}{e\eta_{y,s}}\right].
 \label{eq:tc_detection}
\end{equation}
Here $e>0$ is the elementary charge and $\eta_{j,s}$ denotes
readout efficiency, distinct from the pre-bias signal transmission
$\eta_y$. Optical transmissions after the inverse OPO can be
included in these efficiencies.
A constant background is removed electronically and the remaining
time record is analyzed at the axion frequencies. This operation
assumes many signal cycles in the observation time; subtraction or
servo control must not remove the frequency band being searched.
DC subtraction does not remove the background counting noise or
prevent photodiode saturation. Both detectors must tolerate their individual carrier
powers. A lossless polarization splitter with imperfect extinction
still preserves the collected total; unequal losses and unequal
responsivities must be calibrated.

At $1064\,\mathrm{nm}$ and equal $\eta_d=0.8$, the responsivity is
$\mathcal R=0.686539\,\mathrm{A\,W^{-1}}$. The summed photocurrent
therefore has coefficient
$0.173999\,\mathrm{A\,rad^{-1}}$ and, for the example coupling above,
signal amplitude $9.3385\times10^{-15}\,\mathrm A$.
A known weak polarization rotation inserted before the inverse OPO
calibrates this gain. Reversing $\theta_b$ verifies the predicted
sign reversal of the linear response.

Alternatively, a small time-dependent bias
$\theta_b(t)=\theta_m\cos\Omega_m t$ gives, to leading order,
\begin{equation}
 \delta P_\Sigma(t)=2P_{\rm ext}d_2\theta_m
               \psi_{\rm ext}(t)\cos\Omega_m t.
 \label{eq:tc_modulated_bias}
\end{equation}
For a sinusoidal axion signal, this produces sidebands at
$\Omega_m\pm\Omega_a$. More generally the response is proportional
to $\sin[2\theta_m\cos\Omega_m t]$. Its first harmonic has amplitude
$2J_1(2\theta_m)$, where $J_1$ is the Bessel function of the first
kind. Multiplication by $2\cos\Omega_m t$ followed by low-pass
filtering therefore gives coefficient
$2J_1(2\theta_m)P_{\rm ext}d_2$. This requires a signal band below
$\Omega_m$, rejection of the higher mixing harmonics, and a flat
inverse-OPO and detection response over the detected sidebands.
The bias lies outside the cavity, so it does not require cavity
resonance at $\Omega_m\pm\Omega_a$. For $\theta_m=0.01$ it gives
$5.06862\times10^{-3}\,\mathrm{W\,rad^{-1}}$.
This option requires its own cyclostationary noise calculation;
the static-bias sensitivity below must not be assigned to it.

\subsection{Quantum noise and sensitivity}

We compute a bright-field limit for a static $\pi/4$ bias, zero
polarization splitting and phase-matched ideal OPOs. Both OPO
responses are assumed flat over the plotted band, and all unused
cavity ports carry independent vacuum. Technical noise is set to
zero to define an optical limit. The additional $y$-path loss is
vacuum-coupled and occurs before the bias. Internal OPO loss and static
pump-phase offsets are included in Sec.~\ref{subsec:tc_robustness}.

Let $\Phi_{\rm ext}=P_{\rm ext}/(\hbar\omega_0)$,
$k_2=e^{-r_2}$, and define
\begin{align}
 u_b&=k_2^2\cos^2\theta_b+\sin^2\theta_b,\\
 v_b&=(1-k_2^2)\sin\theta_b\cos\theta_b.
\end{align}
At resonance the amplitude-quadrature spectrum of the extracted
bright field, in vacuum units, is
\begin{align}
 V_{X,\rm ext}(\Omega)&=1+|h_x(\Omega)|^2(e^{2r_1}-1),\\
 h_x(\Omega)&=\frac{T_x e^{i\Omega\ell/c}}
                     {1-q_x e^{i\Omega L_{\rm rt}/c}}.
 \label{eq:tc_carrier_spectrum}
\end{align}
Here $h_x(-\Omega)^*=h_x(\Omega)$ because the carrier cavity is
resonant. The unity term includes all unused vacuum ports.
$V_{X,\rm ext}$ is evaluated \emph{before} the inverse OPO and is
distinct from the post-OPO matrix $V_x$ in
Eq.~\eqref{eq:bright_covariance}. The orthogonal input has vacuum
variance, including after the specified signal loss.

To derive the count noise, define amplitude fluctuations
$\delta X_s=\delta a_{x,s}+\delta a_{x,s}^\dagger$ and
$\delta X_{y,s}=\delta a_{y,s}+\delta a_{y,s}^\dagger$ before the
bias. Use the one-sided continuous-field vacuum normalization
$S_{XX}^{\rm vac}=2$. At zero signal and zero pump-phase error,
linearizing the summed outgoing flux gives
\begin{equation}
 \delta\hat\Phi_{\Sigma,s}
 =C_s\big[u_b\delta\hat X_s+v_b\delta\hat X_{y,s}\big].
 \label{eq:tc_flux_fluctuations}
\end{equation}
The coefficients follow by applying the same bias and inverse-OPO
map to both the carrier and its fluctuations. Thus the output
polarizations cannot be assigned independent Poisson noises.
For equal detector efficiency $\eta_d$, division of the measured
count rate by $\eta_d$ adds the loss-vacuum PSD
$2\Phi_{\rm ext}(1-\eta_d)u_b/\eta_d$ to the summed flux.
Referring the total noise to the rotation with
Eq.~\eqref{eq:tc_power_gain} gives
\begin{equation}
 S_{\psi,\Sigma}^{\rm OPO}(f)=
 \frac{2[u_b^2V_{X,\rm ext}(2\pi f)+v_b^2+
              (1-\eta_d)u_b/\eta_d]}
 {\Phi_{\rm ext}d_2^2\sin^2(2\theta_b)}.
 \label{eq:tc_noise_psd}
\end{equation}
Terms independent of the bright photon flux, such as spontaneous
intensity noise integrated over the OPO bandwidth, are neglected
in this bright-field limit. The formula assumes independent
circulation channels with identical responses; correlated seed or
pump noise belongs in an additional technical covariance.

\begin{figure*}[t]
 \centering
 \includegraphics[width=0.92\textwidth]{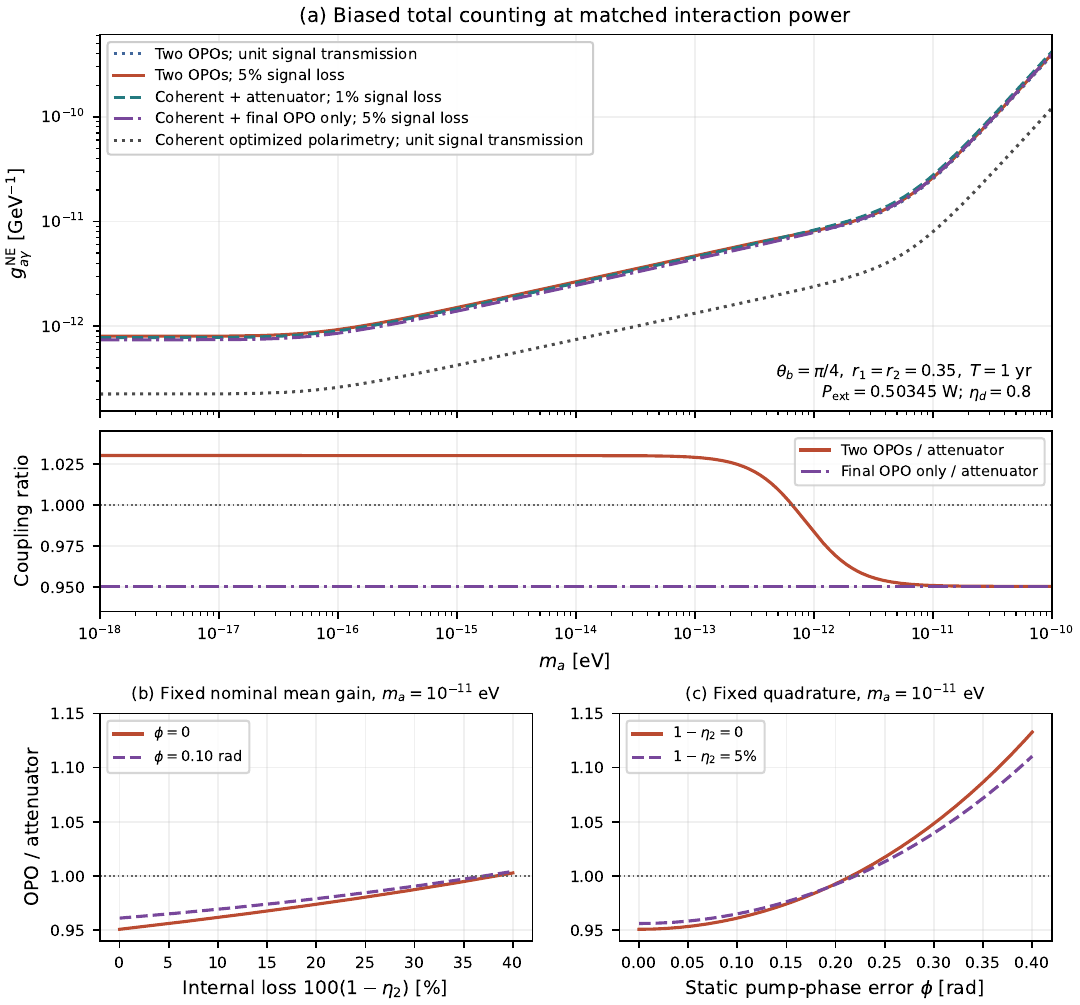}
 \caption{(a) Static-bias total-count sensitivities and their ratios
to coherent total counting with selective attenuation. All cases match
the interaction carrier power and use the same four-port response,
Maxwell line, one-year exposure and $A/\sigma_A=1$ convention. The
nominal total-count curves also match the bias and mean output powers;
their additional signal losses are 5\% for the OPO readouts and 1\%
for the attenuator. The coherent-source/final-OPO curve sets
$V_{X,\rm ext}=1$ and has a coupling ratio of $0.9502$ throughout the
band. The two-OPO curve approaches it above the bright-cavity linewidth.
Optimized coherent polarimetry, with unit signal transmission, is shown
for reference. (b),(c) Two-OPO robustness at $m_a=10^{-11}\,\mathrm{eV}$
under distributed internal loss and a static pump-phase offset,
using Eqs.~\eqref{eq:tc_loss_channel}--\eqref{eq:tc_loss_psd}.
The phase-matched mean deamplification is held fixed in the loss scan;
the quadrature and passive comparator are not retuned for phase error.
Technical phase fluctuations are assessed separately in the text.
Polarization splitting remains zero and the OPO responses are assumed
flat over the displayed band.}

 \label{fig:tc_sensitivity}
\end{figure*}

For a controlled comparison in Fig.~\ref{fig:tc_sensitivity} at fixed interaction carrier power, mean extracted carrier, bias angle, and residual output power (neglecting spontaneous photons in this bright-field limit, as exact single-mode accounting is treated separately), we evaluate Eq.~\eqref{eq:tc_noise_psd} against three reference configurations: (i) a coherent carrier followed by the same bias and a passive $x$-polarization attenuator of power transmission $k_2^2$, which shares the same leading mean response and yields
\begin{equation}
	S_{\psi,\Sigma}^{\rm att}=
	\frac{2u_b}{\eta_d\Phi_{\rm ext}d_2^2\sin^2(2\theta_b)};
	\label{eq:tc_attenuator_noise}
\end{equation}
(ii) a coherent source at the same interaction carrier power followed only by the final OPO, which retains the two-OPO signal coefficient and follows from Eq.~\eqref{eq:tc_noise_psd} by setting $V_{X,\rm ext}=1$; and (iii) optimally phased coherent linear polarimetry, for which $S_\psi^{\rm pol}=1/(2\eta_d\Phi_{\rm ext})$ in this normalization. Because the active and passive total-count readouts share the same response kernel $K_g$ [Eq.~\eqref{eq:tc_coupling_response}] up to their respective pre-bias signal transmissions $\eta_{y,\rm OPO}$ and $\eta_{y,\rm att}$, achieving an OPO sensitivity advantage over the selective attenuator across a locally flat axion line requires
\begin{equation}
	\frac{u_b^2V_{X,\rm ext}+v_b^2+(1-\eta_d)u_b/\eta_d}
	{\eta_{y,\rm OPO}}
	<\frac{u_b}{\eta_d\eta_{y,\rm att}}.
	\label{eq:tc_advantage_condition}
\end{equation}
With equal signal transmission this reduces to
$u_b^2(V_{X,\rm ext}-1)<k_2^2d_2\cos^2\theta_b$.
It identifies how bright-mode excess noise competes with the
vacuum noise of passive attenuation. For a nonuniform line the
information integrals, rather than this pointwise inequality,
must be compared.

In the following information integral, $T\equiv T_{\rm int}$ is the
observation time and $p_a(f;m_a)$ denotes the normalized Maxwellian
line $p_m(f)$ of Eq.~\eqref{eq:maxwell_frequency_line}.
Define the amplitude response per coupling by
\begin{equation}
 K_g(f;m_a)=\frac{\sqrt{\eta_y}}{h_x}
       \frac{\sqrt{2\rho_a}}{2m_a}H_\chi(2\pi f),
 \label{eq:tc_coupling_response}
\end{equation}
with consistent natural units. Numerical couplings in
$\mathrm{GeV}^{-1}$ require the corresponding $10^{-9}$ conversion
to $\mathrm{eV}^{-1}$. The Gaussian stochastic-power convention
used in the main sensitivity analysis gives, in the resolved-line
limit,
\begin{equation}
 (g_{a\gamma}^{\rm NE})^{-4}
 =\frac{T}{4}\int_0^\infty
 \frac{|K_g(f;m_a)|^4p_a(f;m_a)^2}
      {[S_{\psi,\Sigma}(f)]^2}\,df.
 \label{eq:tc_information}
\end{equation}
We use the same finite-time correction as the main analysis for
an unresolved line. Explicitly, for the unboosted Maxwell example,
$p_a=2\sqrt{x}e^{-x}/(\sqrt{\pi}\beta_a)$ for
$x=(f-f_0)/\beta_a\geq0$, $f_0=m_a/(2\pi\hbar)$,
$\beta_a=f_0v_0^2/2$, and $v_0=10^{-3}$
in units of $c$. Multiplying the right-hand side by
$z/(\sqrt{1+z^2}+1)$, $z=2\pi\beta_a T$, is exact when the optical
response and noise are locally constant across the line, with
$f_0T\gg1$ so that the carrier oscillation can be averaged.

For $T=1$ yr, the low-frequency optical-limit plateaus are
$7.83\times10^{-13}\,\mathrm{GeV}^{-1}$ for ideal OPO total counts,
$8.03\times10^{-13}\,\mathrm{GeV}^{-1}$ with 5\% signal loss,
and $7.80\times10^{-13}\,\mathrm{GeV}^{-1}$ for coherent total
counts with 1\% signal loss. Coherent optimized polarimetry gives
$2.26\times10^{-13}\,\mathrm{GeV}^{-1}$ under these idealized
conditions. The coherent-source/final-OPO case gives
$7.41\times10^{-13}\,\mathrm{GeV}^{-1}$ with 5\% signal loss.
The plot stops at
$10^{-10}\,\mathrm{eV}$ ($24.2$ kHz), and assumes sufficient OPO,
photodetector and acquisition bandwidth throughout this range.
It does not extrapolate this readout to the MHz-scale cavity features.

The low-frequency rotation noise is
$\sqrt{S_{\psi,\Sigma}^{\rm OPO}}=1.67\times10^{-9}\,
\mathrm{rad}/\sqrt{\mathrm{Hz}}$.
Bias-angle fluctuations are indistinguishable from an added rotation
in this readout, so their ASD must be below this value to remain
below the optical noise. Relative extracted-carrier power noise
couples with coefficient $u_b/[d_2\sin(2\theta_b)]$; its ASD must
therefore be below approximately
$1.12\times10^{-9}/\sqrt{\mathrm{Hz}}$ at low frequencies if it
is not independently monitored and rejected. These are spectral requirements for a future implementation,
particularly in the coherent low-mass band.

\subsection{Robustness to internal loss and pump-phase error}
\label{subsec:tc_robustness}

We propagate the carrier, both signal sidebands and the full quadrature
covariance through an explicit Gaussian model of loss inside the final
OPO. The loss is uniform along the parametric interaction and couples
to independent vacuum. Its unsqueezed intensity survival is $\eta_2$,
distinct from the extra signal-path transmission $\eta_y$.
Appendix~\ref{subsec:tc_loss_model} gives the channel and its added noise.
To isolate the noise cost at matched nominal output power, the
phase-matched mean deamplification is held at $e^{-r_2}$ by setting the
integrated squeeze parameter to
$r_*=r_2+\tfrac12\ln\eta_2$. This adjustment is possible for
$\eta_2\geq e^{-2r_2}$. We then introduce a static phase offset $\phi$
without retuning the signal quadrature or the passive comparator.
The resulting mean-gain change is included, not treated as noise alone.
This is a specified, frequency-flat internal-loss test, rather than a
characterization of a particular resonant OPO.

At $m_a=10^{-11}\,\mathrm{eV}$ ($f_0=2.42\,\mathrm{kHz}$), the
nominal coupling ratio is $R_0=0.95075$. Table~\ref{tab:tc_robustness}
shows that 5\% internal loss leaves a 4.39\% benefit, and that combining
this loss with $\phi=0.10$ rad leaves a 3.50\% benefit.
Figure~\ref{fig:tc_sensitivity}(b),(c) shows the continuous dependence.
With zero internal loss, the static phase offset erases the advantage
at $|\phi|\simeq0.216$ rad in this model. At low frequencies the two-OPO ratio already exceeds unity, and
these imperfections do not improve it.

\begin{table}[t]
\centering
\caption{Coupling ratio $g_{\rm OPO}^{\rm NE}/g_{\rm att}^{\rm NE}$
from the full line integral at $m_a=10^{-11}\,\mathrm{eV}$, with the
same comparator as Fig.~\ref{fig:tc_sensitivity}. Internal loss refers
to the final OPO; the 5\%/1\% signal-path losses remain fixed.}
\label{tab:tc_robustness}
\begin{tabular}{cccc}
\toprule
$1-\eta_2$ & $\phi$ [rad] & Coupling ratio & Reduction [\%]\\
\midrule
0 & 0 & 0.95075 & 4.93\\
0.01 & 0 & 0.95180 & 4.82\\
0.05 & 0 & 0.95610 & 4.39\\
0.10 & 0 & 0.96171 & 3.83\\
0 & 0.05 & 0.95331 & 4.67\\
0 & 0.10 & 0.96104 & 3.90\\
0.05 & 0.10 & 0.96497 & 3.50\\
\bottomrule
\end{tabular}
\end{table}

The allowable increase in additive noise is modest.
If the OPO signal transfer and passive comparator are unchanged, a
fractional OPO noise-PSD increase $\delta$ that is constant across
the axion line gives
\begin{equation}
 R(\delta)=R_0\sqrt{1+\delta},\qquad
 \delta_{\rm erase}=R_0^{-2}-1.
 \label{eq:tc_noise_margin}
\end{equation}
For the high-frequency ratio $R_0=0.95019$, this is
$\delta_{\rm erase}=0.1076$, or approximately 11\% of the nominal
OPO PSD. Internal loss generally changes signal and noise together,
which is why it was propagated explicitly above.

A static phase tolerance is not a phase-noise specification. For
$\eta_2=1$, a small fluctuation $\delta\phi(t)$ about an offset $\phi_0$
adds an equivalent rotation $k_\phi\delta\phi(t)$, with
\begin{equation}
 k_\phi=\frac{\sinh(2r_2)\sin\phi_0}
 {2[1-\cosh(2r_2)+\sinh(2r_2)\cos\phi_0]}
 \quad(\theta_b=\pi/4).
 \label{eq:tc_phase_noise_gain}
\end{equation}
At $\phi_0=0.10$ rad, $k_\phi=0.0758$ and the remaining PSD margin
at $m_a=10^{-11}\,\mathrm{eV}$ is 8.27\%. If this entire margin is
allocated to uncorrelated pump-phase noise, the local requirement is
$\sqrt{S_\phi}\lesssim5.9\times10^{-9}\,
\mathrm{rad}/\sqrt{\mathrm{Hz}}$. Other noise sources reduce this
allowance. The first-order coefficient vanishes at $\phi_0=0$;
second-order phase fluctuations must still be assessed for an actual
phase-noise spectrum. Polarization splitting and finite OPO bandwidth
are not varied in this total-count test.

\subsection{Role of the first OPO}

With a coherent source at the same interaction carrier power and only
the final OPO, the ratio to the attenuator is $0.95018$ throughout
the displayed band. The two-OPO ratio instead falls from $1.030$ at
low frequency to the same value as the bright-cavity excess noise
is filtered out. Thus the new curve directly attributes the observed
benefit to the final active readout, while the first OPO prepares a
different probe covariance. This readout comparison is separate from
the prospective polarimetric sensitivity region in Eq.~\eqref{eq:prospective_subset} and
respects the information bound in Eq.~\eqref{eq:rotation_qfi_bound}.

\section{Conclusions}

We have developed a four-port ring-cavity polarimetric scheme,
including a downstream electro-optic modulator (EOM) arrangement,
for detecting axion-induced polarization rotation. A central new
result is that a calibrated static polarization bias before the final
optical parametric oscillator (OPO) turns the otherwise quadratic
summed-photon-number response into a linear signal. For our
representative four-port configuration, a $45^\circ$ bias produces a
change of approximately $0.253$ milliwatts in the summed output
optical power for each milliradian of small polarization rotation
in the extracted field. This demonstrates quantitatively how the
axion-induced rotation can be converted into a first-order optical
power signal. Our calculation also accounts for the exact cavity
extraction response, finite axion coherence, correlations between
output polarizations, bright-mode fluctuations, and detection loss.

Our four-port polarimetric sensitivity projection provides a further
important result: under the stated optical and noise assumptions,
it offers the potential to probe a previously unexplored low-mass
region of axion--photon parameter space that is not excluded by the
constraints considered, including the published NuSTAR M82 limits
\cite{NingSafdi2025}. Specifically, we identify a prospective window
at axion masses between $10^{-18}$ and $10^{-16}\,\mathrm{eV}$ and
couplings from $3.2\times10^{-13}\,\mathrm{GeV}^{-1}$ to below
$6.0\times10^{-13}\,\mathrm{GeV}^{-1}$. This window lies above the
projected noise-equivalent sensitivity threshold even when the
assumed 20\% technical-noise power-spectral-density allowance is
included. The stated reach uses the $A/\sigma_A=1$ stochastic-power
sensitivity convention and the adopted local axion density; it is
not a confidence-level exclusion forecast, nor does it establish
that constraints not included in our comparison are absent.

Separately, we demonstrate an additional original readout result:
at matched interaction and extracted carrier powers, OPO-based
total counting reduces the noise-equivalent axion--photon coupling
by approximately 5\% relative to the specified selective-attenuation
comparator in the favorable optical band above the bright-cavity
linewidth. In our distributed-loss test, 5\% internal loss leaves
a 4.4\% improvement at an axion mass of $10^{-11}\,\mathrm{eV}$,
while combining this loss with a static pump-phase offset of
$0.10$ rad leaves a 3.5\% improvement. The advantage is modest:
an approximately 11\% additive increase in the nominal OPO noise
power spectral density would eliminate it if the signal transfer
remained unchanged. We further show that a coherent source followed
only by the final OPO retains the high-frequency readout benefit
without the first OPO's low-frequency excess noise, identifying
active readout conversion, rather than probe preparation, as the
origin of this improvement.

Optimized coherent polarimetry remains more sensitive in the
benchmark considered. Thus the prospective access to a region not
excluded by the constraints examined and the approximately 5\%
improvement in the matched total-count comparison are distinct
findings. Together, our results establish the potential of the
proposed four-port polarimetric architecture for future axion
searches while identifying the optical loss, pump-phase stability,
technical-noise, and detector-bandwidth requirements that must be
verified experimentally.

\section*{ACKNOWLEDGMENTS}
A. C. and G. P. acknowledge partial financial support from MIUR.
The authors acknowledge support from the Istituto Nazionale di Fisica
Nucleare (INFN) through the Iniziativa Specifica QGSKY.

\section*{Data and code availability}
This is a theoretical study. The optical, halo and statistical
assumptions are specified in the text; no measured instrumental-noise
spectra are used. Published bounds retain the assumptions of their
original analyses.

\appendix
\section{Canonical derivation of the off-resonant Hamiltonian}
\label{app:birefringence}

For completeness, we derive the canonical form underlying Sec.~\ref{sec:hamiltonian_rwa}.  With metric $\eta_{\mu\nu}=\mathrm{diag}(1,-1,-1,-1)$ and $F_{\mu\nu}\widetilde F^{\mu\nu}=-4\mathbf E\cdot\mathbf B$, the semiclassical photon Lagrangian density is
\begin{equation}
    \mathcal L_\gamma
    =\frac12\left(\mathbf E^2-\mathbf B^2\right)
    +g_{a\gamma}a(t)\mathbf E\cdot\mathbf B.
    \label{eq:L_semiclassical_appendix}
\end{equation}
In Coulomb gauge, $A^0=0$ and $\boldsymbol\nabla\cdot\mathbf A=0$, this becomes
\begin{equation}
    L_\gamma
    =\int d^3x\left[
    \frac12\dot{\mathbf A}^{\,2}
    -\frac12\mathbf B^2
    -g_{a\gamma}a(t)\dot{\mathbf A}\cdot\mathbf B
    \right].
    \label{eq:L_coulomb_appendix}
\end{equation}
The canonical momentum is
\begin{equation}
    \boldsymbol\Pi
    =\frac{\delta L_\gamma}{\delta\dot{\mathbf A}}
    =\dot{\mathbf A}-g_{a\gamma}a(t)\mathbf B,
    \label{eq:Pi_appendix}
\end{equation}
and the exact photon Hamiltonian in these canonical variables is
\begin{equation}
    H_\gamma
    =\int d^3x\left[
    \frac12\left(\boldsymbol\Pi+g_{a\gamma}a\mathbf B\right)^2
    +\frac12\mathbf B^2
    \right].
    \label{eq:Hcanonical_appendix}
\end{equation}
Thus the frequently used identification $H_{\rm int}=-L_{\rm int}$ is not exact in the original canonical variables because the interaction depends on $\dot{\mathbf A}$.

For a spatially homogeneous axion field, define the magnetic helicity functional
\begin{equation}
    K[\mathbf A]=\int d^3x\,\mathbf A\cdot\mathbf B.
\end{equation}
Under periodic boundary conditions, perfectly reflecting idealized boundaries, or after separating the open-system boundary coupling from the bulk dynamics,
\begin{equation}
    \frac{dK}{dt}=2\int d^3x\,\dot{\mathbf A}\cdot\mathbf B.
    \label{eq:helicity_identity_appendix}
\end{equation}
Consequently,
\begin{equation}
    -g_{a\gamma}a\int d^3x\,\dot{\mathbf A}\cdot\mathbf B
    =\frac{g_{a\gamma}}{2}\dot a\,K
    -\frac{d}{dt}\left(\frac{g_{a\gamma}}{2}aK\right).
\end{equation}
Dropping the total derivative gives the canonically equivalent Lagrangian stated in Sec.~\ref{sec:hamiltonian_rwa}.  Its canonical momentum is $\mathbf P=\dot{\mathbf A}$, and its Hamiltonian is Eq.~\eqref{eq:hamiltonian_prime_main}.  The two Hamiltonians, Eqs.~\eqref{eq:Hcanonical_appendix} and \eqref{eq:hamiltonian_prime_main}, must not be added: they represent the same dynamics in canonical variables related by a time-dependent transformation.

At a fixed time, the transverse field is expanded as
\begin{equation}
    \hat{\mathbf A}(\mathbf x)
    =\sum_{h=\pm1}\int\frac{d^3k}{(2\pi)^3}
    \frac{1}{\sqrt{2k}}
    \left[
    \boldsymbol\epsilon_h(\mathbf k)\hat c_{\mathbf kh}e^{i\mathbf k\cdot\mathbf x}
    +\mathrm{H.c.}
    \right],
    \label{eq:A_helicity_appendix}
\end{equation}
where $i\mathbf k\times\boldsymbol\epsilon_h=hk\boldsymbol\epsilon_h$.  Taking the curl and integrating over space yields Eq.~\eqref{eq:AB_helicity_main}.  Substitution into Eq.~\eqref{eq:hamiltonian_prime_main} gives, to first order in $g_{a\gamma}$,
\begin{align}
    \hat H_\gamma(t)
    ={}&\sum_h\int\frac{d^3k}{(2\pi)^3}k\hat N_{\mathbf kh}
    -\frac{g_{a\gamma}}{2}\dot a(t)
    \sum_hh\int\frac{d^3k}{(2\pi)^3}
    \nonumber\\
    &\times\left[
    \hat N_{\mathbf kh}
    +\frac12\hat c_{\mathbf kh}\hat c_{-\mathbf kh}
    +\frac12\hat c_{\mathbf kh}^{\dagger}\hat c_{-\mathbf kh}^{\dagger}
    \right].
    \label{eq:Hfull_helicity_appendix}
\end{align}
The number-conserving term produces the opposite helicity shifts, while the remaining terms generate pair annihilation and creation.  In the interaction picture the pair operators oscillate as $e^{\mp2ikt}$.  For an axion field with frequencies concentrated near $m_a$, their near-resonant phase is $e^{\mp i(2k-m_a)t}$.  Therefore, sufficiently far from $m_a=2k$, coarse graining over times long compared with $k^{-1}$ removes these terms and gives Eq.~\eqref{eq:H_RWA_main}.  In the optical mass range considered here, $m_a/k\lesssim10^{-6}$, so the separation from the parametric resonance is parametrically large.

The linear-to-circular transformation is an ordinary unitary change of basis in the two-dimensional transverse polarization space of the same massless photon field.  In Nambu form it has vanishing anomalous block,
\begin{equation}
    \begin{pmatrix}
        \hat{\boldsymbol c}_{\rm circ}\\
        \hat{\boldsymbol c}_{\rm circ}^{\dagger}
    \end{pmatrix}
    =
    \begin{pmatrix}
        W&0\\
        0&W^*
    \end{pmatrix}
    \begin{pmatrix}
        \hat{\boldsymbol c}_{\rm lin}\\
        \hat{\boldsymbol c}_{\rm lin}^{\dagger}
    \end{pmatrix},
    \qquad W^\dagger W=\mathbb I.
    \label{eq:nambu_passive_appendix}
\end{equation}
It therefore leaves the vacuum and total photon number invariant.  This passive basis change should not be confused with the non-number-conserving Bogoliubov dynamics generated by the pair terms in Eq.~\eqref{eq:Hfull_helicity_appendix}.

\section{Bright-mode closure and orthogonal-signal escape}
\label{app:bright_mode_closure}

The single-mode OPO acts only on the $x$ polarization. With $\mu=\cosh r$ and $\nu=\sinh r$,
\begin{align}
    S_x^\dagger\hat a_xS_x&=\mu\hat a_x+\nu\hat a_x^\dagger,
    &S_x^\dagger\hat a_yS_x&=\hat a_y,
    \\
    S_x^{-1\dagger}\hat a_xS_x^{-1}&=\mu\hat a_x-\nu\hat a_x^\dagger,
    &S_x^{-1\dagger}\hat a_yS_x^{-1}&=\hat a_y.
\end{align}
For the passive rotation
\begin{align}
    U_a^\dagger\hat a_xU_a&=\hat a_x\cos\psi_a-\hat a_y\sin\psi_a,
    \\
    U_a^\dagger\hat a_yU_a&=\hat a_y\cos\psi_a+\hat a_x\sin\psi_a,
\end{align}

composition gives Eq.~\eqref{eq:output_ops}. The identity at zero
rotation applies only to the ideal lossless sequence. The additional
vacuum inputs of an open cavity are treated in
Appendix~\ref{app:input_output}.

For $\alpha=|\alpha|e^{i\vartheta}$,
\begin{equation}
    \mu\alpha+\nu\alpha^*
    =|\alpha|\left(e^r\cos\vartheta+i e^{-r}\sin\vartheta\right),
\end{equation}
which yields the phase-dependent mean amplitude stated in Sec.~\ref{sec:su11_direct_readout}. The maximum signal gain $e^r$ therefore requires phase locking of the seed and OPO pump. This transformation is distinct from the axion-induced Bogoliubov pair sector discarded by the RWA: the OPO deliberately mixes $\hat a_x$ and $\hat a_x^\dagger$, whereas the axion evolution retained in the RWA is a passive number-conserving rotation between $x$ and $y$.

\subsection{Total-count fluctuations and pump work}

For the phase-matched seed let $s=\sin\psi_a$, $c=\cos\psi_a$ and
$\nu=\sinh r$. In quadratures with unit vacuum variance, the
lossless map is $M=S_x^{-1}R(\psi_a)S_x$, and the covariance is
$V=MM^T$. For two modes with quadrature displacement $\bm d$,
$\operatorname{Var}(\hat N_{\rm tot})=
[\operatorname{Tr}(V^2)-4]/8+\bm d^TV\bm d/4$.
The resulting exact variance is
\begin{align}
 \operatorname{Var}(\hat N_{\rm tot})={}&
 n_0\big[(c^2+e^{2r}s^2)^2+4\nu^2c^2s^2\big]
 \nonumber\\
 &+4\nu^2s^2(1+\nu^2s^2).
 \label{eq:total_photon_variance}
\end{align}
It reduces to $n_0$ at the null. In the bright-seed limit, the
mean-excess statistic therefore has
$\Delta N_{\rm tot}/\sqrt{\operatorname{Var}(\hat N_{\rm tot})}
\simeq(e^{2r}-1)\sqrt{n_0}\,\psi_a^2$.
For independent repetitions this is a quadratic-intensity readout,
with rotation reach proportional to the inverse fourth root of the
accumulated seed photon number at fixed $r$. This statement is about
the bright-background mean-count statistic; it does not apply to an
empty-input dark-count measurement or replace the full Fisher bound.

The ideal pump work can be separated into
$W_1=\hbar\omega_0(n_S-n_0)$ and
$W_2=\hbar\omega_0(N_{{\rm tot,out}}-n_S)$, so that
$W_1+W_2=\hbar\omega_0\Delta N_{\rm tot}$. The second term can be
negative. Energy is conserved when the pump fields are included.
For unequal detection efficiencies, a measured count sum is not the
optical total. With a common efficiency $\eta_d$, binomial thinning
gives $\langle N_d\rangle=\eta_dN_{{\rm tot,out}}$ and
$\operatorname{Var}(N_d)=\eta_d^2\operatorname{Var}(\hat N_{\rm tot})
+\eta_d(1-\eta_d)N_{{\rm tot,out}}$.
The actual figure uses a dump for the bright output; measuring
this total would require monitoring that output as well.

\section{Propagation reversal and detection-port transport}
\label{app:cw_ccw_sign}

For two free waves propagating in opposite directions, the relation
between helicity and a fixed laboratory transverse basis is reversed.
Up to an overall rotation-sign convention, their Jones rotations are
\(U_s=\exp(s\psi_aJ)\), with \(s=\pm1\),
where $\psi_a$ is the single-pass angle of
Eq.~\eqref{eq:psi_general_main}. Thus the off-diagonal conversion
changes sign when both waves are analyzed in that same basis.
This statement concerns antiparallel free waves, not the parallel
beams leaving different cavity ports.

In Fig.~\ref{fig:apparatus_closed_eom}, let the physical mirror
positions be $(0,d),(\ell,d),(\ell,0),(0,0)$ for M1--M4.
The propagation orders and ports are
\begin{align}
 \mathrm{CW}:&\quad(\mathrm{M1,M2,M3,M4}),
       &&\mathrm{in=M1},\quad\mathrm{out=M2},\nonumber\\
 \mathrm{CCW}:&\quad(\mathrm{M4,M3,M2,M1}),
       &&\mathrm{in=M4},\quad\mathrm{out=M3}.
 \label{eq:physical_port_orders}
\end{align}
In each travel ordering the segment sequence is
$(\ell,d,\ell,d)$. Take identical mirror matrices
$R_k=\operatorname{diag}(-r_x,r_y e^{-i\delta_{\rm rt}/4})$
in the transported local frames, identical transmission maps, and
equal reference amplitudes and EOM phases in the parallel-output
bases. The cyclic equations in
Eqs.~\eqref{eq:cyclic_carrier} and \eqref{eq:cyclic_signal}
are then identical in the two travel orderings, including the direct
long segment leading to extraction. Both the carrier transfer and
the axion conversion at the output are equal. Subsequent identical
OPO and analyzer maps give equal calibrated signal transfers, as stated in Sec.~\ref{sec:ring_cavity}.
The analyzer redirects reference and signal together; a common
output phase does not change their interference.

This construction specifies one symmetric realization, not the sign
for arbitrary coatings or output optics. For general mirror maps,
solve the two ordered propagation problems separately and evaluate
their detector-basis interference as in Sec.~\ref{sec:ring_cavity}.
An electronic or EOM phase reversal can change a channel's reported
sign, but also changes its response to instrumental polarization
disturbances. Signal preservation and noise rejection must therefore
be tested using their respective calibrated response vectors.

\section{Two-polarization input--output and covariance model}
\label{app:input_output}

\subsection{Mixing phase and measured quadrature}

The intracavity operators $c_x,c_y$ describe the nearly orthogonal
polarization modes of the open cavity. In the linear-polarization
convention of the main text, a real rotation rate
$\Omega_a^{\rm rot}(t)$ is generated by
\begin{equation}
 H_{\rm mix}=i\hbar\Omega_a^{\rm rot}(t)
                   (c_y^\dagger c_x-c_x^\dagger c_y).
 \label{eq:real_rotation_hamiltonian}
\end{equation}
It gives $\dot c_y|_a=\Omega_a^{\rm rot}c_x$ and
$\dot c_x|_a=-\Omega_a^{\rm rot}c_y$.
This mixing rate is distinct from the axion oscillation frequency
$\Omega_a$. Its effective cavity value must be matched to the
ordered geometric source, rather than assigned from the round-trip
length alone.

With the back-action on the strong carrier neglected to first order,
a local resonance model is
\begin{align}
 \dot c_x={}&-(\gamma_x+i\Delta_x)c_x
      +\sum_p\sqrt{2\gamma_{x,p}}\,a_{x,p}^{\rm in}
      +\sqrt{2\gamma_{x,\ell}}\,n_x,\nonumber\\
 \dot c_y={}&-(\gamma_y+i\Delta_y)c_y
      +[\Omega_a^{\rm rot}(t)+\kappa_{\rm mir}(t)]c_x\nonumber\\
      &+\sum_p\sqrt{2\gamma_{y,p}}\,a_{y,p}^{\rm in}
      +\sqrt{2\gamma_{y,\ell}}\,n_y.
 \label{eq:input_output_intracavity}
\end{align}
The index $p$ runs over all coupling ports,
$\gamma_j=\sum_p\gamma_{j,p}+\gamma_{j,\ell}$ is the amplitude
half-linewidth, and $n_j$ is the absorption/scattering bath.
$\Delta_y-\Delta_x$ is the static polarization splitting.
The mirror-mixing coefficient $\kappa_{\rm mir}$ is generally complex
and is not identified with a real axion rotation.
Retaining an $i\Omega_a^{\rm rot}c_x$ source would instead require
rephasing the signal mode and the optical readout reference together.

For $c_x=\bar c_x+\delta c_x$ and Fourier convention $e^{-i\Omega t}$,
\begin{align}
 c_y(\Omega)={}&\chi_y(\Omega)\Big[\bar c_x
      (\Omega_a^{\rm rot}+\kappa_{\rm mir})(\Omega)\nonumber\\
      &+\sum_p\sqrt{2\gamma_{y,p}}\,a_{y,p}^{\rm in}(\Omega)
      +\sqrt{2\gamma_{y,\ell}}\,n_y(\Omega)\Big],\nonumber\\
 \chi_y(\Omega)={}&[\gamma_y+i(\Delta_y-\Omega)]^{-1}.
 \label{eq:cavity_susceptibility}
\end{align}
At a distinct output port $o$,
$a_j^{\rm out}=a_{j,o}^{\rm in}-\sqrt{2\gamma_{j,o}}c_j$;
the sign is a port-phase convention. If $H_y(\Omega)$ is the
complex signal-field transfer from a real drive, its measured
quadrature transfer at reference phase $\theta$ is
\begin{equation}
 H_\theta(\Omega)=\frac12\left[
 e^{-i\theta}H_y(\Omega)+e^{i\theta}H_y^*(-\Omega)\right].
 \label{eq:sideband_quadrature}
\end{equation}
Both optical sidebands are necessary. A modulus of one sideband
alone does not specify the detected quadrature.

For the time-domain Nambu vector
\[
 \bm v=(a_x,a_y,a_x^\dagger,a_y^\dagger)^T,
\]
the external sequence is
\begin{equation}
 \bm v_{\rm det}=\sqrt{\eta_{\rm ro}}\,
 {\cal R}_m(t){\cal S}_{x,2}^{-1}\bm v_{\rm out}
 +\sqrt{1-\eta_{\rm ro}}\,\bm v_{\rm vac},
 \label{eq:external_eom_detection}
\end{equation}
where ${\cal R}_m$ is the passive EOM rotation. At zero residual
pump-phase error,
\begin{equation}
 {\cal S}_{x,2}^{-1}(r_2)=
 \begin{pmatrix}
 \cosh r_2&0&-\sinh r_2&0\\
 0&1&0&0\\
 -\sinh r_2&0&\cosh r_2&0\\
 0&0&0&1
 \end{pmatrix}.
 \label{eq:inverse_opo_nambu}
\end{equation}
The EOM is outside the cavity and is absent from its propagation
map. A fixed polarization retarder or an equivalent optical phase
control may be included in ${\cal R}_m$ to select $\theta$.

\subsection{Exact cyclic equations for the specified four ports}

For the model in Sec.~\ref{subsec:four_port_study}, take
\begin{equation}
 T_j=1-\sqrt{q_j},\quad r_j=q_j^{1/4},\quad
 R_k=\operatorname{diag}(-r_x,r_ye^{-i\delta_{\rm rt}/4}),
 \label{eq:four_port_mirrors}
\end{equation}
where $j=x,y$, $k=0,\ldots,3$ follows each physical travel order,
and $\delta_{\rm rt}=\Delta_yT_{\rm rt}$.
At high finesse $\delta_{\rm rt}\simeq
(\Delta_y/\gamma_y)(1-q_y)$, the convention used numerically.
For carrier phase gauged into the mirror maps,
\begin{equation}
 \bar{\bm C}_k=R_k\bar{\bm C}_{k-1}+t_k\bm\beta_k,
 \qquad t_k=\operatorname{diag}(\sqrt{T_x},\sqrt{T_y}),
 \label{eq:cyclic_carrier}
\end{equation}
with cyclic indices; $\bm\beta_k$ is the coherent input-field
amplitude vector and is nonzero only for the driven port $k=0$.
Let $z_j=e^{i\Omega\tau_j}$, where the flight times in either
travel ordering are $(\ell,d,\ell,d)/c$. The coherent sideband obeys
\begin{equation}
 \begin{split}
 \delta\bm C_k=R_k\Big[&z_{k-1}\delta\bm C_{k-1}\\
 &+\frac{g_{a\gamma}a_\Omega}{2}(1-z_{k-1})J\bar{\bm C}_{k-1}
 \Big].
 \end{split}
 \label{eq:cyclic_signal}
\end{equation}
The quantum fluctuation equations additionally have the input term
$t_k\bm v_k$ at every mirror, where $\bm v_k$ denotes the
independent vacuum-field vector incident at that port.
Transmission at $k=1$ gives the
coherent output signal
\begin{equation}
 \begin{split}
 Y_s=\sqrt{T_y}\,\bm e_y^T\Big[&z_0\delta\bm C_0\\
 &+\frac{g_{a\gamma}a_\Omega}{2}(1-z_0)J\bar{\bm C}_0\Big].
 \end{split}
 \label{eq:four_port_extraction}
\end{equation}
For the quantum output, the reflected vacuum incident at the
extraction port is included with the complementary reflection
coefficient required by its unitary beam-splitter map. All four
vacuum ports are independent. Solving the two cyclic systems
retains both repeated conversions and the direct segment to the
output. In the lossless passive $y$ subsystem the complete scattering
map preserves unit vacuum variance.

\subsection{Closed transfer including the extraction segment}
\label{subsec:closed_four_port}

The cyclic equations can be solved without a uniform-carrier
approximation. Define $z_\ell=\exp(i\Omega\ell/c)$,
$z_d=\exp(i\Omega d/c)$ and
$\varrho_y=r_y\exp(-i\delta_{\rm rt}/4)$.
With the drive immediately after mirror $k=0$ and extraction at
$k=1$, write $Y_s=(g_{a\gamma}a_\Omega/2)\beta_sH_{4p}$.
Then
\begin{align}
 H_{4p}(\Omega)&=\frac{\sqrt{T_xT_y}}{1-q_x}
 \frac{{\cal N}(\Omega)}{1-\varrho_y^4(z_\ell z_d)^2},
 \label{eq:closed_four_port_transfer}\\
 {\cal N}(\Omega)&=(1-z_\ell)(1+\varrho_y^2r_x^2z_\ell z_d)
 \nonumber\\
 &\quad-\varrho_yr_xz_\ell(1-z_d)
             (r_x^2+\varrho_y^2z_\ell z_d).
 \label{eq:closed_four_port_numerator}
\end{align}
The terms proportional to $1-z_\ell$ include the direct source
between the input and extraction mirrors. The remaining terms
carry the signs and weights acquired on the other segments.
The symmetric physical port orders in
Appendix~\ref{app:cw_ccw_sign} give this same transfer for both
calibrated output records, so the signal-preserving combination is
their sum. Changing a port or a mirror map requires solving its
own cyclic system.

The carrier transfer is $h_x=T_x/(1-q_x)$, and
$H_{4p}(\Omega)=-i\Omega{\cal L}_{4p}/c+O(\Omega^2)$, where
\begin{equation}
 {\cal L}_{4p}=\frac{\sqrt{T_xT_y}}{(1-q_x)(1-\varrho_y^4)}
 \left[(1+r_x^2\varrho_y^2)\ell
 -r_x\varrho_y(r_x^2+\varrho_y^2)d\right].
 \label{eq:four_port_effective_length}
\end{equation}
At zero splitting this is $4.27166\times10^4\,\mathrm m$ for
the stated mirrors. The common fixed signal-quadrature angle is
$\theta=\arg{\cal L}_{4p}$, with the static carrier phase separately
calibrated. At finite frequency Eq.~\eqref{eq:sideband_quadrature}
is applied to $H_{4p}$, not to its modulus.

Finite port transmissions are retained in both the numerator and
denominator. In particular, the scalar-model masses
$m_n=2\pi n\hbar/T_{\rm rt}$ need not be zeros of this transfer.
For example, at $m_1=1.23984\times10^{-7}\,\mathrm{eV}$ and zero
splitting, $|H_{4p}|=1.55187$; it is nonzero even for a
monochromatic drive. The axion linewidth must subsequently be
integrated over this actual response.

\subsection{Carrier, signal gain and readout noise}

Write the incident amplitude in each channel as
$\beta_s=e^{r_1}\alpha_s$, with
$\sum_s|\alpha_s|^2=P_0/(\hbar\omega_0)$ in flux units.
For the resonant example $h_x$ is real; after the second OPO,
\begin{equation}
 B_s=h_x\beta_s(\cosh r_2-e^{i\phi}\sinh r_2).
 \label{eq:residual_carrier}
\end{equation}
In quadratures with unit vacuum variance, define
\begin{equation}
 A_2=\begin{pmatrix}
 \cosh r_2-\sinh r_2\cos\phi&-\sinh r_2\sin\phi\\
 -\sinh r_2\sin\phi&\cosh r_2+\sinh r_2\cos\phi
 \end{pmatrix}.
 \label{eq:inverse_opo_quadratures}
\end{equation}
The quasistatic bright-mode covariance is
\begin{equation}
 V_x=A_2\left[
 |h_x|^2\begin{pmatrix}e^{2r_1}&0\\0&e^{-2r_1}\end{pmatrix}
 +(1-|h_x|^2)I\right]A_2^T.
 \label{eq:bright_covariance}
\end{equation}
For $r_1=r_2=0.35$ and $\phi=0$, the variance parallel to the
carrier is $V_{x,\parallel}=0.62244$. The passive signal channel at
the null remains $V_y^{\rm pre}=1$, with static cross-polarization
mixing set to zero in this example. Post-cavity signal power loss
$\ell_y$ gives
\begin{align}
 G_{\rm sig}&=e^{r_1}\sqrt{1-\ell_y},\nonumber\\
 \overline P_{\rm ref,det}&=
 \eta_{\rm ro}\hbar\omega_0\sum_s|B_s|^2
                      \langle\sin^2\psi_m\rangle_m.
 \label{eq:gain_reference_independent}
\end{align}
Thus $G_{\rm sig}$, $V_y^{\rm pre}$ and the reference power are
separate outputs of the model. The reference is not assigned the
seed power after a lossy cavity.

To include finite analyzer extinction, set $\vartheta=\psi_m(t)$ and
$p=\sin^2\vartheta+\epsilon\cos^2\vartheta$. The detector collects
the transmitted $y$ field and a fraction $\epsilon$ of the orthogonal
bright power. For channel $s$ the null rate and its linearized
noise intensity are
\begin{align}
 \lambda_s^{(0)}&=\eta_{\rm ro}|B_s|^2p,\nonumber\\
 D_s&=\eta_{\rm ro}|B_s|^2
     \left[p+\eta_{\rm ro}(V_{x,\parallel}-1)p^2\right].
 \label{eq:finite_extinction_count_noise}
\end{align}
The linear signal includes the interference
$\operatorname{Re}(B_s^*Y_s)$, multiplied by
$(1-\epsilon)\sin\vartheta\cos\vartheta$. The $p^2$ term retains the
bright-mode correlations after the EOM and detection vacuum; treating
the two collected polarizations as unrelated Poisson beams would
discard it. At zero leakage the bright-mode correction is below
$4\times10^{-5}$ for the chosen modulation depth.

In this quasistatic linearized limit the information for the
signal-rotation amplitude $\theta_{\rm sig}$ is
$\sum_s\int(\partial_{\theta_{\rm sig}}\lambda_s)^2/D_s\,dt$; it reduces to
the Poisson information of Sec.~\ref{sec:linear_lockin} for coherent inputs.
For $r_1=r_2=r$ and a real signal quadrature before the residual
pump-phase error, the modulation factor evaluated numerically is
\begin{equation}
 F_m=\left\langle
 \frac{(1-\epsilon)^2\sin^2\vartheta\cos^2\vartheta}
 {p+\eta_{\rm ro}(V_{x,\parallel}-1)p^2}
 \right\rangle_m.
 \label{eq:finite_extinction_modulation_factor}
\end{equation}
In the following penalty, $H_y$ is the signal-field transfer with
the stated splitting and $H_{y,0}$ its reference value at zero
splitting, evaluated with the same optical normalization.
The sensitivity penalty, after correcting the static splitting
quadrature but retaining the residual pump-phase error, is
\begin{equation}
 \mathcal P=
 \frac{1}{\sqrt{1-\ell_y}\,|H_y/H_{y,0}|\,
              |\cos\arg B_s|\sqrt{F_m}}.
 \label{eq:combined_readout_penalty}
\end{equation}
This prescription yields Table~\ref{tab:parameter_penalties}.
It assumes a bright carrier and approximately constant gain and
covariance over the relevant quasistatic band. Frequency-dependent
noise requires the full sideband covariance, not this static matrix.

For the equal-resource coherent comparison, the selective attenuator
gives the same $B_s=h_x\alpha_s$ at zero phase error and the same
signal field, with $V_x=I$. The difference in leading sensitivity
vanishes as $\psi_0\to0$; at $\psi_0=10^{-2}$ the calculated
bright-covariance correction is below $10^{-5}$ in the ideal
sensitivity ratio. Additional signal loss and technical noise must
be assessed separately using the loss and noise comparison in Sec.~\ref{sec:power_accounting}.

\subsection{Distributed internal loss in the final OPO}
\label{subsec:tc_loss_model}

For the robustness test, write quadratures in the order
$(X_x,P_x,X_y,P_y)$, with vacuum covariance $I_4$.
Uniform vacuum damping and squeezing along a normalized interaction
coordinate $0\leq z\leq1$ define the Gaussian channel on $x$,
$\boldsymbol q_{x,\rm out}=M\boldsymbol q_{x,\rm in}
+\boldsymbol\xi$. With $\ell_2=-\ln\eta_2$,
$r_*=r_2-\ell_2/2$ and $O=R(\phi/2)$, where $R$ is the real
two-dimensional rotation in Eq.~\eqref{eq:tc_rotation},
\begin{align}
 M&=\sqrt{\eta_2}\,O
       \operatorname{diag}(e^{-r_*},e^{r_*})O^T,\nonumber\\
 N&=\langle\boldsymbol\xi\boldsymbol\xi^T\rangle_{\rm sym}
     =O\operatorname{diag}(n_-,n_+)O^T,\nonumber\\
 n_{\mp}&=\ell_2\frac{1-e^{-(\ell_2\pm2r_*)}}
                         {\ell_2\pm2r_*}.
 \label{eq:tc_loss_channel}
\end{align}
The removable zero in each quotient is taken by continuity.
These noise terms follow from integrating the vacuum injected along
the interaction, $N=\ell_2\int_0^1 e^{-\ell_2z}
O\operatorname{diag}(e^{-2r_*z},e^{2r_*z})O^Tdz$.
At $\phi=0$, the carrier transmission is
$\sqrt{\eta_2}e^{-r_*}=e^{-r_2}$, so the nominal mean output power
is maintained during the loss scan. At $\eta_2=1$, $N=0$ and $M$
reduces to Eq.~\eqref{eq:inverse_opo_quadratures}.

Let $R_b=R(\theta_b)\otimes I_2$ act on the two polarizations and
set $T_2=(M\oplus I_2)R_b$, $N_2=N\oplus0_2$.
The carrier and real signal directions before these operations are
$\boldsymbol e_x=(1,0,0,0)^T$ and
$\boldsymbol e_y=(0,0,1,0)^T$.
Define $\boldsymbol b=T_2\boldsymbol e_x$,
$\boldsymbol w=T_2^T\boldsymbol b$, $u=\boldsymbol b^T\boldsymbol b$
and $\mathcal L=2\boldsymbol b^TT_2\boldsymbol e_y$.
The extracted-field covariance is
\begin{align}
 V_0(f)&=\operatorname{diag}(V_X,V_P,1,1),\nonumber\\
 V_{X,P}(f)&=1+|h_x(2\pi f)|^2(e^{\pm2r_1}-1).
 \label{eq:tc_loss_input_cov}
\end{align}
For the coherent-source/final-OPO case $V_0=I_4$ at the same carrier
flux. The rotation-referred summed-count PSD is
\begin{equation}
 S_{\psi,\Sigma}(f)=
 \frac{2\big[\boldsymbol w^TV_0(f)\boldsymbol w+
 \boldsymbol b^TN_2\boldsymbol b+(1-\eta_d)u/\eta_d\big]}
 {\Phi_{\rm ext}\mathcal L^2}.
 \label{eq:tc_loss_psd}
\end{equation}
This projection includes output-polarization correlations and all
loss vacua. At $\eta_2=1$, $\phi=0$ it reproduces
Eq.~\eqref{eq:tc_noise_psd}. Zero splitting and the fixed quadrature
$\chi=0$ ensure that the axion mean is in the real signal direction;
its temporal phase is retained by $H_\chi$ in
Eq.~\eqref{eq:tc_coupling_response}. Thus both sidebands and the changed
gain $\mathcal L$ enter the same information integral as the noise.
The analytic loss covariance, the mean-field derivative and the full
four-quadrature projection provide independent consistency checks.

\subsection{Coherent leakage and a spectral error bound}
\label{subsec:coherent_leakage_counts}

The wide-band calculation extends the preceding static model with
a coherent parasitic field $Y_{\rm par}/B=i\xi$ in the signal
polarization. This field is specified after the quadrature selector
and before the EOM. Its phase need not follow a fluctuating carrier.
In the calibrated frame where the static $B$ is real, set
$\vartheta=\psi_0\cos\Omega_mt$ and define
\begin{align}
 p_0&=\sin^2\vartheta+\epsilon\cos^2\vartheta,\nonumber\\
 n&=\cos^2\vartheta+\epsilon\sin^2\vartheta,\nonumber\\
 j&=(1-\epsilon)\sin\vartheta\cos\vartheta,\nonumber\\
 p_\xi&=p_0+\xi^2n.
 \label{eq:coherent_leakage_factors}
\end{align}
For a small real signal $Y/B=q$, the null count rate and its
derivative are $\lambda_s^{(0)}=\eta_{\rm ro}|B_s|^2p_\xi$ and
$\partial_q\lambda_s=2\eta_{\rm ro}|B_s|^2j$.
Thus the leading Poisson factor is
\begin{equation}
 F_m^{(P)}=\left\langle\frac{j^2}{p_\xi}\right\rangle_m.
 \label{eq:coherent_leakage_modulation}
\end{equation}
It includes the parasitic photon background rather than treating
that field only as a technical-noise coupling.

For a quasistatic bright covariance $V_x$ expressed in this
carrier frame, the noise intensity is
\begin{equation}
 D_s=\eta_{\rm ro}|B_s|^2
 \left[p_\xi+\eta_{\rm ro}\bm t^T(V_x-I)\bm t\right],
 \qquad \bm t=(p_0,\xi j)^T.
 \label{eq:coherent_leakage_covariance}
\end{equation}
At $\xi=0$ this reduces to the static model above. For a spectral
calculation $V_x$ is generally frequency dependent, so its static
value cannot simply be used as white noise over the plotted band.
Instead, bounded OPO gains $|r_1(f)|,|r_2(f)|\leq0.35$ and passive
cavity propagation give $\|V_x(f)-I\|\leq e^{2(r_1+r_2)}-1$.
Together with the modulation projections this bounds the relative
correction to the weighted count-noise PSD by
\begin{equation}
 \varepsilon_V\leq\eta_{\rm ro}
 [e^{2(r_1+r_2)}-1]
 (\sin^2\psi_0+\epsilon+\xi^2)<2.47\times10^{-4}.
 \label{eq:bright_spectral_error_bound}
\end{equation}
Accordingly, using Eq.~\eqref{eq:coherent_leakage_modulation} in
the broad-band curves changes the noise-equivalent coupling by less
than $1.3\times10^{-4}$ fractionally under these gain assumptions.
This bound concerns quantum bright-mode fluctuations; it does not
bound pump, polarization or electronic technical spectra.

The count information assumes that the modulation and its signal
sidebands are resolved and that the time-dependent likelihood weight
is implemented. Many modulation and signal cycles, or equivalent
phase averaging, give Eq.~\eqref{eq:output_shot_psd}. An ordinary
cosine lock-in has a different modulation efficiency. Finite detector
bandwidth and the realized filter must be included when translating
this optical limit into an instrumental prediction.

\subsection{Finite-time stochastic information}
\label{subsec:finite_time_information}

For $A=g_{a\gamma}^2$ the covariance per unit $A$ is
\begin{equation}
 C_a(\tau)=\frac12\int_0^\infty
 |{\cal K}(f;m_a)|^2p_m(f)\cos(2\pi f\tau)\,df.
 \label{eq:axion_covariance_per_power}
\end{equation}
In white one-sided noise $S_q$, the weak-signal Fisher information
on a record of length $T$ is
\begin{equation}
 {\cal I}_A=\frac4{S_q^2}\int_0^T(T-\tau)C_a(\tau)^2\,d\tau.
 \label{eq:finite_covariance_information}
\end{equation}
For a transfer constant over the line, the normalized Maxwell
envelope has squared modulus
$[1+(2\pi\beta_a\tau)^2]^{-3/2}$. Averaging the fast carrier
oscillation, justified by $f_0T\gg1$, reduces the remaining integral to
\begin{equation}
 \int_0^T\frac{T-\tau}{[1+(2\pi\beta_a\tau)^2]^{3/2}}\,d\tau
 =\frac{T^2}{\sqrt{1+(2\pi\beta_a T)^2}+1}.
 \label{eq:maxwell_finite_integral}
\end{equation}
This gives Eq.~\eqref{eq:finite_time_power_information}. The
long-time limit recovers Eq.~\eqref{eq:stochastic_information}.

The numerical broad-band curve evaluates the full frequency
integral and multiplies its information by
$x_T/[\sqrt{1+x_T^2}+1]$, where $x_T=2\pi\beta_a T$.
This matching is exact for a line-flat transfer after fast-cycle
averaging and tends to unity in the long-time limit. In the present
parameter range the coherence transition is far below optical
features; near those features $x_T\gg1$. It is therefore a
controlled interpolation for this example, not an exact formula
for an arbitrary finite-time, colored-noise response. Such a response
requires the full covariance operator and inverse noise covariance.

The line integral is evaluated with 128 generalized Gauss--Laguerre
nodes after the substitution $u=2(f-f_0)/\beta_a$ and checked with
256 nodes. Independent cyclic-matrix and truncated-Fock-space
checks verify the transfer and photon moments. The total-count
extensions additionally check the mean derivative, quadrature
projection and vacuum covariance of the dissipative OPO channel.

\makeatletter
\let\auto@bib@innerbib\@empty
\makeatother
\linespread{0.91}\selectfont

\end{document}